\documentclass[a4paper, 12 pt] {article}

\usepackage{amsmath}
\usepackage{amsfonts}
\usepackage{amsthm}
\usepackage{amscd} 
\usepackage{amssymb}

\newtheorem{thm}{Theorem}[section]
\newtheorem{lem}{Lemma}[section]

\newtheorem{defn}{Definition}[section]

\usepackage{graphicx}
\usepackage{epsfig}
\usepackage{graphics}

\newcommand\Div{ \,\text{div}\, } 
\newcommand\Curl{\,\text{\bf curl}\,}
\renewcommand\Re{\,\text{Re}\,}
\renewcommand\Im{\,\text{Im}\,}

\newlength{\intwidth}

\newcommand{\half}{{\textstyle \frac{1}{2}}}
\newcommand{\fourth}{{\textstyle \frac{1}{4}}}

\newcommand{\fpi}{{\textstyle \frac{1}{4\pi}}}

\newcommand\smallfrac[2]{{\textstyle \frac{#1}{#2}}}

\newcommand{\Evec}{{\bf E}}
\newcommand{\mvec}{{\bf m}}
\newcommand{\nvec}{{\bf n}}
\newcommand{\pvec}{{\bf c}}
\newcommand{\Nvec}{{\bf N}}

\newcommand{\Cvec}{{\bf C}}
\newcommand{\Rvec}{{\bf R}}

\newcommand\grad{{\bf \nabla}}
\newcommand\gradn{{\bf \nabla}_n}
\newcommand\gradt{{\bf \nabla}_t}
\newcommand\pvecn{{\bf c}_n}
\newcommand\pvect{{\bf c}_t}
\newcommand\avec{{\bf a}}

\newcommand\cvec{{\bf c}}
\newcommand\Fvec{{\bf F}}
\newcommand\tvec{{\bf t}}

\newcommand\vvec{{\bf v}}

\newcommand\Fhat{{  \hat F}}

\newcommand\jhat{{\bf  \hat j}}
\newcommand\khat{{\bf  \hat k}}
\newcommand\lhat{{\bf  \hat l}}

\newcommand\xhat{{\bf \hat x}}
\newcommand\yhat{{\bf \hat y}}
\newcommand\zhat{{\bf \hat z}}
\newcommand\chit{{\tilde \chi}}
\newcommand\inff{\text{inf}}

\newcommand\sgn{\text{sgn}\,}

\newcommand{\rvec}{{\bf r}}
\newcommand{\svec}{{\bf s}}

\newcommand{\nuvec}{{\boldsymbol \nu}}

\newcommand{\Acal}{{\cal HC}}
\newcommand{\Ccal}{{\cal C}}
\newcommand{\Ecal}{{\cal E}}
\newcommand{\Gcal}{{\cal G}}
\newcommand{\Pcal}{{\cal P}}
\newcommand{\Ncal}{{\cal N}}

\newcommand{\Rr}{{\mathbb R}}
\newcommand{\Cc}{{\mathbb C}}
\newcommand{\Zz}{{\mathbb Z}}

\newcommand\tbar{{\bar t}}
\newcommand\wbar{{\bar w}}

\newcommand\Abar{{\tilde A}}

\newcommand{\bvec}{{\bf b}}
\newcommand{\gvec}{{\bf g}}

\newcommand{\gammavec}{{\boldsymbol \gamma}}
\newcommand{\Gammavec}{{\boldsymbol \Gamma}}

\numberwithin{equation}{section}
\begin{document}



\title{Energies of $S^2$-valued harmonic maps on polyhedra with tangent
  boundary conditions}

\author{A Majumdar$^{\dag\ \ddag}$, 
JM Robbins$^\dag$
\& M Zyskin$^\dag$ 
\thanks{a.majumdar@bristol.ac.uk, j.robbins@bristol.ac.uk, m.zyskin@bristol.ac.uk}\\
School of Mathematics\\ University of Bristol, University Walk, 
Bristol BS8 1TW, UK\\
and\\
Hewlett-Packard Laboratories,\\
 Filton Road, Stoke Gifford, Bristol BS12 6QZ, UK}

\thispagestyle{empty}

\maketitle

\begin{abstract}
  A unit-vector field $\nvec:P \rightarrow S^2$ on a convex polyhedron
  $P \subset \Rr^3$ satisfies tangent boundary conditions if, on each
  face of $P$, $\nvec$ takes values tangent to that face.  Tangent
  unit-vector fields are necessarily discontinuous at the vertices of
  $P$.  We consider fields which are continuous elsewhere.  We derive
  a lower bound $E^-_P(h)$ for the infimum Dirichlet energy
  $E^\inff_P(h)$ for such tangent unit-vector fields of arbitrary homotopy
  type $h$.  $E^-_P(h)$ is expressed as a weighted sum of minimal
  connections, one for each sector of a natural partition of $S^2$
  induced by $P$.  For $P$ a rectangular prism, we derive an upper
  bound for $E^\inff_P(h)$ whose ratio to the lower bound may be bounded
  independently of $h$.  The problem is motivated 
by models of
  nematic liquid crystals in polyhedral geometries.  Our results
  improve and extend several previous results.

\end{abstract}


\newpage

\section{Introduction}\label{sec:intro}

$S^2$-valued harmonic maps on three-dimensional domains with holes
were studied in a well known paper by  Brezis, Coron and Lieb
\cite{brecorlie}. As a simple representative example, consider the domain
$\Omega = \Rr^3 - \{\rvec^1,\ldots,\rvec^n\}$ (for which the holes are
points), and let $\nvec:\Omega\rightarrow S^2$ denote a unit-vector
field on $\Omega$.  For $\grad \nvec$ square-integrable, we define the
Dirichlet energy of $\nvec$ to be
\begin{equation}\label{eq:Omega energy}
   E(\nvec) = \int_\Omega(\grad \nvec)^2 dV.
\end{equation}
Continuous unit-vector fields on $\Omega$ may be classified up to
homotopy by their degrees, $d = (d^1,\ldots,d^n)\in \Zz^n$, on spheres
about each of the excluded points (the restriction of $\nvec$ to such
a sphere may be regarded as a map from $S^2$ into itself).  In order
that $\nvec$ have finite energy, we must have that
\begin{equation}
  \label{eq:d_sum}
  \sum_j d^j = 0.
\end{equation}
Let $\Ccal_\Omega(d)$ denote the homotopy class of continuous
unit-vector fields with degrees $d$ satisfying (\ref{eq:d_sum}), and
let $\Acal_\Omega(d)$ denote the elements of $\Ccal_\Omega(d)$ with
finite Dirichlet energy.  Let $E^\inff_\Omega(d)$ denote the infimum of the energy over
$\Acal_\Omega(d)$,
\begin{equation}
  \label{eq:E_h_BCL}
  E^\inff_\Omega(d) = \inf_{\nvec \in \Acal_\Omega(d)} E(\nvec).
\end{equation}

It turns out that $E^\inff_\Omega(d)$ is just $8\pi$ times the
length of a {\it minimal connection} on $\Omega$.  We recall the
definition of a minimal connection.
Given two $m$-tuples of points in $\Rr^3$, $\Pcal = (\avec^1,\ldots, \avec^m)$ and
$\Ncal = (\bvec^1,\ldots,\bvec^m)$ (whose points need not be distinct), a
connection is a pairing $(\avec^j,\bvec^{\pi(j)})$ of points in $\Pcal$ and $\Ncal$,
specified here in terms of a permutation $\pi \in S_m$ ($S_m$ denotes the
symmetric group).  The length of
a connection is the sum of the distances between the paired points,
and a minimal connection is a connection with minimum length.  Let
\begin{equation}
  \label{eq:minimal_connection}
  L(\Pcal,\Ncal) = \min_{\pi \in S_m} \sum_{j  = 1}^m |\avec^j - \bvec^{\pi(j)}|
\end{equation}
denote the length of a minimal connection between $\Pcal$ and $\Ncal$.  Let $|d| = \half
  \sum_j |d^j|$.
Then
\begin{thm}\label{thm:0}{\rm \cite{brecorlie}}
The infimum $E^\inff_\Omega(d)$ of the Dirichlet energy of continuous
unit-vector fields on the domain $\Omega = \Rr^3 - \{\rvec^1,\ldots,\rvec^n\}$
of given degrees $d^j$ about the excluded points $\rvec^j$
is given by 
  \begin{equation}
    \label{eq:BCL-result}
    E^\inff_\Omega(d) = 8\pi L(\Pcal(d),\Ncal(d)),
  \end{equation}
  where $\Pcal(d)$ is the $|d|$-tuple of excluded points of positive degree, 
with $\rvec^j$ included $d^j$ times, and 
$N(d)$ is the $|d|$-tuple of excluded points of negative degree,
with $\rvec^k$ included $|d^k|$ times.
\end{thm}

In this paper we consider a natural variant of this problem which emerges from
a boundary-value problem of some physical and technological interest;
the domain is taken to be
a polyhedron on which $\nvec$ is required to satisfy {\it tangent boundary conditions}.
Let $P$ denote a convex bounded
polyhedron in $\Rr^3$, including the interior of the polyhedron but excluding
its vertices.  Let $\nvec:P\rightarrow S^2$ denote a unit-vector field
on $P$.  We say that $\nvec$ satisfies tangent boundary conditions, or
is tangent, if, on each face of $P$, $\nvec$ takes values tangent to
that face.  (It is clear that this condition could not be satisfied at
the vertices of the polyhedron, which belong to three or more faces). 

One motivation for the
problem comes from liquid crystals applications, in which $\nvec$ describes
the mean local orientation of a nematic liquid crystal, and the
Dirichlet energy (\ref{eq:energy}) coincides with the elastic or Frank-Oseen
energy in the so-called one-constant approximation (see, eg,
\cite{degennes, virga, klemanlav, stewart}).  Polyhedral
cells have been proposed as a mechanism for engendering bistability --
they may support two nematic configurations with distinct optical
properties, both of which are local minima of the elastic energy
 \cite{jones, newtonspiller, kg2002}.
In many cases of interest the orientation at
interfaces is well described by tangent boundary conditions, and
low-energy local
minimisers appear to have different topologies.  
We also  remark that harmonic maps between Riemannian polyhedra have
been studied by Gromov and Shoen \cite{gs} and Eells and Flugede
\cite{eellsfug}, in particular in cases where the target manifold
has nonpositive curvature.

Here we will restrict our attention to continuous tangent unit-vector
fields on $P$. (Let us note that, while nematic orientation is, in general,
described by a director, or $RP^2$-valued field, a continuous director
field on a simply connected domain such as $P$ can be lifted to a
continuous unit-vector field.)
Continuous tangent unit-vector fields on $P$ can be partitioned into
homotopy classes $\Ccal_P(h)$ labeled by a complete set of homotopy
invariants denoted collectively by $h$.  A full account of this
classification is given in
\cite{rz2003} (see also \cite{z2005}).  Below we reprise the results
we need here.  General discussions  of topological defects in liquid
crystals are given in \cite{mermin, kleman, klemanlav}.

For $\grad \nvec$ square integrable on $P$,
let 
\begin{equation}
  \label{eq:energy}
  E(\nvec) = \int_P (\grad \nvec)^2 \,dV
\end{equation}
denote its Dirichlet energy.
Let $\Acal_P(h)$ denote the elements of $ \Ccal_P(h)$ with
finite Dirichlet energy, and let
\begin{equation}
  \label{eq:infimum}
  E^\inff_P(h) = \inf_{\nvec \in \Acal_P(h)} E(\nvec)
\end{equation}
denote the infimum of the energy over $\Acal_P(h)$.  Our first result
(Theorem~\ref{thm:1} below) is a lower bound for $E^\inff_P(h)$.  This
is expressed is terms of certain homotopy invariants called wrapping
numbers, which we now define.  Let $f$ denote the number of faces of
$P$, $F^c$ the $c$th face of $P$, and $\Fvec^c$ the outward normal on
$F^c$, where $1 \le c \le f$.  For each face we consider the great
circle on $S^2$ containing the unit-vectors $\svec$ tangent to it, ie
$\Fvec^c\cdot \svec = 0$.  These $f$ (not necessarily distinct) great
circles partition $S^2$ into open spherical polygons, which we call
{\it sectors}.
The sectors are characterised by $\sgn (\Fvec^c\cdot \svec)$.  We write
\begin{equation}
  \label{eq:S_sigma}
  S^\sigma = \{\svec \in S^2 \,|\, \sgn (\svec\cdot \Fvec^c) = \sigma^c\},
\end{equation}
where $\sigma = (\sigma^1,\ldots, \sigma^f)$ is an $f$-tuple of signs.
It should be noted that most of the $S^\sigma$'s are empty;
indeed, reckoning based on Euler's formula for polygonal partitions
of the sphere ($|$vertices$|$ - $|$edges$|$ + $|$faces$|$ = 2) shows that there
are at most (and generically) $f^2-f+2$ nonempty sectors.

Next, let $C^a$ denote a smooth surface in $P$ which separates the
vertex $\vvec^a$ from the others.  For definiteness, take $C^a$ to be oriented so that
$\vvec^a$ lies on the positive side of $C^a$.  We call $C^a$ a {\it
  cleaved surface}.  Given $\nvec \in \Ccal_P(h)$, let $\nvec^a$
denote its restriction to $C^a$. The wrapping number $w^{a\sigma}$
is the number of times $\nvec^a$ covers $S^\sigma$, counted with
orientation.
For $\nvec$ differentiable, this is given by 
\begin{equation}
  \label{eq:wrapping_def_as_integal}
  w^{a\sigma} = \frac{1} {A^\sigma} \int_{C^a} \,
\nvec^*(\chi^\sigma \omega), 
\end{equation}
where $\omega$ is the area two-form on $S^2$, normalised to
have integral
$4\pi$, $\chi^\sigma$
is the characteristic function of $S^\sigma \subset S^2$, and 
$  A^\sigma = \left| 
\int_{S^2} \chi^\sigma \,\omega 
\right |$
is the   area
of $S^\sigma$.
Alternatively, $w^{a\sigma}$ can be expressed as
the index of a
regular value $\svec \in S^\sigma$ of $\nvec^a$, ie
\begin{equation}
  \label{eq:wrapping numbers}
  w^{a\sigma} = \sum_{\rvec\, | \, \nvec^a(\rvec) = \svec} \sgn \det d\nvec^a(\rvec).
\end{equation}
One can show that
$w^{a\sigma}$ does not depend on the choice of $\svec\in S^\sigma$
nor on the choice of cleaving surface $C^a$, that the definition
(\ref{eq:wrapping_def_as_integal}) 
can be extended to continuous
$\nvec$, and that its value depends only on the homotopy class of
$\nvec$ \cite{rz2003}.  In fact, the wrapping numbers constitute a
complete set of invariants, as is shown in the Appendix.
They are not independent, however.  For
example, continuity in the interior of $P$ (absence of singularities) implies
that, for all $\sigma$,
\begin{equation}
  \label{eq:w sum rule}
  \sum_{a} w^{a\sigma} = 0,
\end{equation}
where the sum is taken over vertices $\vvec^a$.
Continuity on the faces and edges of $P$ implies additional
constraints.  We will say that $h = \{w^{a\sigma}\}$ is an {\it admissible topology} if
it can be realised by some continuous configuration $\nvec: P
\rightarrow S^2$.

\begin{thm}\label{thm:1}
Let $h = \{w^{a\sigma}\}$ be an admissible topology for continuous tangent
unit-vector fields on a polyhedron $P$.  Then
\begin{equation}
  \label{eq:E_-}
  E^\inff_P(h)\ge E^-_P(h) := \sum_\sigma 2 A^\sigma L(\Pcal^\sigma(h), \Ncal^\sigma(h)),
\end{equation}
where $\Pcal^\sigma$
(resp.~$\Ncal^\sigma$) contains the vertices of $P$  for which
$w^{a\sigma}$ is positive (resp.~negative), each such vertex included
with multiplicity 
$|w^{a\sigma}|$.
\end{thm}
\noindent Thus, to each sector $\sigma$ may be associated a constellation of point
defects at the vertices $\vvec^a$ of degrees $w^{a\sigma}$.
The lower bound $E^-_P(h)$ is a sum of the lengths of minimal connections for
these constellations weighted by the areas of the sectors.

Theorem~\ref{thm:1} is proved using arguments similar to those used to
show that $E^\inff_\Omega(d) \ge 8\pi L(\Pcal(d),\Ncal(d))$ in the
proof of Theorem~\ref{thm:0}.
In Theorem~\ref{thm:0}, one obtains an equality for $E^\inff_\Omega(d)$,
rather than just a lower bound, by constructing a sequence  $\nvec^{(j)}$
whose energies approach $8\pi L(\Pcal(d),\Ncal(d))$.  It can be shown
that a subsequence 
$\nvec^{(k)}$ approaches a 
constant away from a minimal connection while
$|\grad \nvec^{(k)}|^2$ approaches a singular measure supported on the
minimal connection \cite{brecorlie}. 
In the present case, tangent boundary conditions preclude such 
a construction; 
$\nvec$ is required to vary across the
faces and, therefore, throughout the interior of $P$.  However, for $P$
a rectangular prism, we can show that $E^-_P(h)$ correctly describes
the dependence of the infimum energy on homotopy type.
\begin{thm}
  Let $P$ denote a rectangular prism with sides of length $L_x \ge L_y
  \ge L_z$ and largest aspect ratio $\kappa = L_x/L_z$.  Then
  \label{thm:2}
  \begin{equation}
    \label{eq:th2}
    E_P^\inff(h) \le C \kappa^3 
E^-_P(h)
  \end{equation}
  for some constant $C$ independent of $h$ and $L_x$, $L_y$, $L_z$.
\end{thm}
\noindent The upper bound of Theorem~\ref{thm:2} is obtained by estimating
the energy of explicitly constructed tangent unit-vector fields which
satisfy the Euler-Lagrange equations near each vertex.

The general form of the Frank-Oseen energy is given by \cite{degennes,
  virga, klemanlav, stewart}
\begin{multline}
  \label{eq:elastic}
  E_{FO}(\nvec) = \int_P \big [ K_1 (\Div \nvec)^2 + K_2 (\nvec
  \cdot \Curl \nvec)^2 +   K_3
  (\nvec\times \Curl \nvec)^2\\ +  K_4 \Div ((\nvec\cdot
  \grad)\nvec - (\Div \nvec) \nvec)\big ]\, dV,
\end{multline}
where the elastic constants $K_j$ are material-dependent.
It is easily shown that tangent boundary conditions imply
that the contribution from the $K_4$-term in (\ref{eq:elastic}) vanishes.  The 
elastic constants $K_1$, $K_2$ and $K_3$ are constrained to be
nonnegative, and the one-constant approximation (\ref{eq:energy})
follows from taking $K_1 = K_2 = K_3 = 1$.  We remark that
Theorems~\ref{thm:1} and \ref{thm:2} imply the following bounds for
the Frank-Oseen energy:
\begin{equation}
  \label{eq:FO_bound}
K_- E^-_P(h)
\le \inf_{\nvec \in \Acal_P(h)} E_{FO}(\nvec) \le C K_+ \kappa^3   E^-_P(h),
\end{equation}
where $K_-$ (resp.~$K_+$) is the smallest (resp.~largest) of the
elastic constants $K_1$, $K_2$ and $K_3$.

Theorems~\ref{thm:1} and \ref{thm:2} improve
and extend  several earlier results.  In \cite{mrz2004a} we
obtained a lower bound for $E^\inff_P(h)$ of the form
\begin{equation}
  \label{eq:LMP_lower_bound}
  2 \max_{\xi^a} \sum_a \xi^a \left(\sum_\sigma  A^\sigma w^{a\sigma}\right),
\end{equation}
where the maximum is taken over $\xi^a$'s such that $|\xi^a - \xi^b|
\le |\vvec^a - \vvec^b|$.  The quantity (\ref{eq:LMP_lower_bound}) is
generally less than the lower bound given by Theorem~\ref{thm:1}, in
particular because it allows for cancellations between wrapping
numbers of opposite sign.  For example, for a regular tetrahedron with
sides of unit length, (\ref{eq:LMP_lower_bound}) gives a lower bound
of $\sum_{a\sigma} A^\sigma w^{a\sigma}$, whereas Theorem~\ref{thm:1}
gives the lower bound $\sum_{a\sigma} A^\sigma |w^{a\sigma}|$.  For a
rectangular prism, Theorem~\ref{thm:2} does not hold if
(\ref{eq:LMP_lower_bound}) is substituted for $E^-_P(h)$.
((\ref{eq:LMP_lower_bound}) can be directly compared to (\ref{eq:9})
below, which gives an equivalent (dual) expression for $E^-_P(h)$.)
A restricted
example of Theorem~\ref{thm:1} was given in \cite{mrz2006} for the
case 
$h$ is a {\it
  reflection-symmetric topology}.  These are  the topologies of configurations
which are invariant under reflections through the midplanes of the
prism.

Results related to Theorem~\ref{thm:2} were obtained for the special case of
reflection-symmetric topologies in \cite{mrz2004b, mrz2006}.  
The constructions and estimates are simpler in this case, and one can show
that the ratio of the upper and lower bounds scales linearly with the
aspect ratio $\kappa$, rather than as $\kappa^3$.  Indeed, for
conformal and anticonformal reflection-symmetric topologies (for which
the wrapping numbers $w^{a\sigma}$ about a given vertex have the same sign), one can
show that the ratio is bounded by $(L_x^2 + L_y^2 + L_z^2)/L_z$.  It
is not clear that for general prism topologies the $\kappa^3$ dependence
in Theorem~\ref{thm:2} is optimal.


An important question is whether within a given homotopy class the
infimum Dirichlet energy is achieved.  The homotopy classes
$\Acal_P(h)$ are not weakly closed with respect to the Sobolev norm,
so it is not automatically the case that the infimum is achieved.
However, while a given $\Acal_P(h)$ may not be weakly closed, it may
still contain a local (smooth) minimiser of the Dirichlet energy.
Indeed, there is some numerical evidence and heuristics to suggest
that, in the case of a rectangular prism, for the simplest topologies
a smooth minimiser always exists, while for others a smooth local
minimiser may or may not exist depending on the aspect ratios
\cite{mrz2004b, majthesis}.  It would be interesting to establish for
which topologies there exist smooth local minimisers, also from the
point of view of device applications.  Such configurations would of
course satisfy the bounds established here.

There is an extensive literature on $S^2$-valued harmonic maps with
fixed (Dirichlet) boundary data;  reviews are given in \cite{hardt97,brezis06}.
Problems related to the one considered here concern liquid crystal
droplets \cite{lavrentovich, virga}, in which one seeks configurations
$\nvec$ on a three-dimensional region $\Omega$ which minimise the
elastic energy \cite{linpoon}.  In case of tangent boundary
conditions, there are necessarily singularities on the surface of
$\Omega$, eg 'boojums' \cite{lavrentovich2}; in a polyhedral domain,
these singularities are pinned at the vertices.  



The remainder of the paper is organised as follows.
Theorem~\ref{thm:1} is proved in Section~\ref{sec: lower bound}, and
Theorem~\ref{thm:2} in Section~\ref{sec: upper bound} modulo two
lemmas concerning the explicit construction of and estimates for the
representative prism configurations (Sections~\ref{sec: pf of lemma 1}
and \ref{sec: pf of lemma 2}).  In the Appendix it is shown that
homotopy classes of continuous tangent unit-vector fields on $P$ are
classified by wrapping numbers.

\section{Lower bound for general polyhedra}\label{sec: lower bound}

\begin{proof}[Proof of Theorem \ref{thm:1}]
In \cite{mrz2004a} we show that smooth $\nvec$'s are dense in
$\Acal_P(h)$ with respect to the Sobolev $W^{(1,2)}$-norm.  Therefore, it
suffices to establish the lower bound (\ref{eq:E_-}) for $\nvec$ smooth.

Let $B_\epsilon(\vvec^a)$ denote the $\epsilon$-ball about $\vvec^a$,
and let 
\begin{equation}
  \label{eq:P_epsilon}
  P_\epsilon = P - \cup_a(B_\epsilon(\vvec^a)\cap P)
\end{equation}
denote the domain obtained by excising these balls from $P$.  Clearly
 \begin{equation}
   \label{eq:1a}
   E(\nvec) = \int_P  (\nabla \nvec)^2
\, dV \ge  \int_{P_\epsilon} (\nabla \nvec)^2
\, dV.
 \end{equation}
Let $\chi^\sigma$ denote the characteristic function of the sector
$S^\sigma\subset S^2$.  It will be useful to introduce
smooth approximations $\chit^\sigma$ to $\chi^\sigma$, such that
$\chit^\sigma$
has support in $S^\sigma$ and satisfies $0 \le \chit^\sigma \le
\chi^\sigma$.  Then
\begin{equation}
  \label{eq:3}
  E(\nvec) \ge \sum_\sigma \int_{P_\epsilon} 
 (\chit^\sigma\circ \nvec)\,  (\nabla
  \nvec)^2\, dV . 
\end{equation}
Using the inequality \cite{brecorlie}
\begin{equation}
  \label{eq:localinequality}
  (\nabla \nvec)^2\ \ge 2 ||\nvec^* \omega||,
\end{equation}
where $\nvec^*\omega$ denotes the pullback of $\omega$ by $\nvec$ and $|| \cdot ||$
denotes the norm on forms induced by the standard metrics on $\Rr^3$ and $S^2$, 
we  get that
\begin{equation}
  \label{eq:1}
  E(\nvec) \ge 2\sum_\sigma \int_{P_\epsilon} 
||\nvec^* (\chit^\sigma\omega)|| \, dV.
\end{equation}

For each $\sigma$, let $\xi^\sigma$ denote a continuous piecewise-differentiable function
on $P$ with 
\begin{equation}
  \label{eq:d_xi}
  ||d\xi^\sigma|| \le 1.
\end{equation}
Then for arbitrary $\avec$, $\bvec$, $\cvec$, we have that
\begin{equation}
  \label{eq:4}
  ||\nvec^* \omega||\,\left|dV(\avec,\bvec,\cvec)\right|  \ge (d\xi^\sigma \wedge \nvec^*
   \omega)(\avec, \bvec, \cvec),
\end{equation}
where $dV$ is here regarded as the Euclidean volume form on $\Rr^3$.
But
\begin{equation}
  \label{eq:6}
d\xi^\sigma\wedge \nvec^* (\chit^\sigma\omega) = d(\xi^\sigma\wedge \nvec^* (\chit^\sigma\omega)),
\end{equation}
since $d(\nvec^*(\chit^\sigma\omega)) = \nvec^* d(\chit^\sigma\omega) = 0$
($\chit^\sigma\omega$ is a  two-form on $S^2$).  Therefore,
\begin{equation}
  \label{eq:5}
   E(\nvec) \ge 2\sum_\sigma \int_{P_\epsilon}
d\left(\xi^\sigma\wedge \nvec^* (\chit^\sigma\omega)\right).
\end{equation}

From Stokes' theorem,  (\ref{eq:5}) implies that
\begin{equation}
  \label{eq:8}
    E(\nvec) 
\ge 2 \sum_\sigma \int_{\partial P_\epsilon}
\xi^\sigma \nvec^*(\chit^\sigma \omega), 
\end{equation}
The boundary of $P_\epsilon$ consists of i) 
the faces of $P$ 
with points in $B_\epsilon(\vvec^a)$ removed, and ii) the
intersections of 
the 
two-spheres $\partial B_\epsilon(\vvec^a)$ with $P$.  The latter, denoted
by $C^a_\epsilon = \partial B_\epsilon(\vvec^a) \cap P$, we call
{\it cleaved surfaces}.
Tangent boundary conditions imply that $\nvec^*\omega$ 
vanishes on the faces of $P$ (since the values of $\nvec$ on a face are
restricted to a great circle in $S^2$).  Therefore, only the cleaved
surfaces contribute to the integral in (\ref{eq:8}). We obtain
\begin{equation}
  \label{eq:8.1}
    E(\nvec) 
\ge 2\sum_\sigma \sum_a \int_{C^a_\epsilon}
\xi^\sigma \nvec^*(\chit^\sigma \omega).
\end{equation}
By the Bounded Convergence Theorem, we can 
replace $\chit^\sigma$ by
$\chi^\sigma$ in 
(\ref{eq:8.1}). Taking 
the limit
$\epsilon \rightarrow 0$, 
we obtain
\begin{equation}
  \label{eq:8a}
    E(\nvec) \ge 
2\sum_\sigma \sum_a \xi^\sigma(\vvec^a) \lim _{\epsilon\rightarrow 0}
\int_{C^a_\epsilon}
\nvec^* (\chi^\sigma \omega).
\end{equation}
From (\ref{eq:wrapping_def_as_integal}), 
the integral over $C^a_\epsilon$  yields $A^\sigma$ times  the
wrapping number $w^{a\sigma}$, which depends only on the homotopy
type of $\nvec$.  Thus,
\begin{equation}
  \label{eq:8b}
    E_P^\inff(h) \ge 
2 \sum_\sigma A^\sigma \sum_a   \xi^\sigma(\vvec^a) w^{a\sigma}.
\end{equation}

The remainder of the argument proceeds as in  \cite{brecorlie}.  We
note that (\ref{eq:8b}) holds for any choice of $\xi^\sigma$'s
consistent with the constraints (\ref{eq:d_xi}).  These constraints
imply that
\begin{equation}
  \label{eq:xi_constraints}
  |\xi^\sigma(\vvec^a) - \xi^\sigma(\vvec^b)| \le |\vvec^a - \vvec^b|.
\end{equation}
Conversely, given any set of values $\xi^{a\sigma}$ for which
$|\xi^{a\sigma} - \xi^{b\sigma}| \le |\vvec^a - \vvec^b|$, we can find
functions $\xi^{\sigma}$ which satisfy the constraints (\ref{eq:d_xi})
and assume these values at the vertices (for example, take
$\xi^\sigma(\rvec) = \max_a (\xi^{a\sigma} - |\rvec - \vvec^a|)$).
Thus, we obtain a lower bound for $E_P(h)$ in terms of the solutions
of a finite number of linear optimisation problems, one for each
sector,
\begin{equation}
  \label{eq:9}
    E_P^\inff(h) \ge 
2 \sum_\sigma A^\sigma  \left (\max_{|\xi^{a\sigma} - \xi^{b\sigma}|
  \le |\vvec^a - \vvec^b|} \sum_a  \xi^{a\sigma} 
w^{a\sigma}\right).
\end{equation}

A simpler characterisation is provided by the dual formulation,
\begin{equation}
  \label{eq:10}
    E^\inff_P(h) \ge 
2 \sum_\sigma  A^\sigma \left(\min_{\Omega^{ab,\sigma}}  \sum_{a,b}
|\vvec^a - \vvec^b| \Omega^{ab,\sigma} 
\right),
\end{equation}
where the $\Omega^{ab,\sigma}$'s are constrained by
\begin{equation}
  \label{eq:11}
\sum_a \Omega^{ab,\sigma} = -w^{b\sigma}, \quad 
\sum_b \Omega^{ab,\sigma} =  w^{a\sigma}.
\end{equation}
Let us fix $\sigma$. Without loss of generality, we can restrict $\Omega^{ab,\sigma}$ to be
nonnegative and equal to zero unless $w^{a \sigma}>0$ and $w^{b
  \sigma}<0$.  Suppose first that the nonzero wrapping numbers are
either $+1$ or $-1$;
by (\ref{eq:w sum rule}) there are an equal number,
$m$ say, of each.  Therefore, the nonvanishing elements of $\Omega^{ab,\sigma}$
may be identified with an $m\times m$
matrix, which we denote by $M$.
(\ref{eq:11})
implies that $M$ is doubly stochastic.
By a theorem of Birkhoff \cite{birkhoff}, $M$ can be expressed as a convex linear
combination of permutation matrices. 
Then the minimum in (\ref{eq:10}) is necessarily achieved at an
extremal point, ie for 
$M$ 
a permutation matrix corresponding to a minimal connection. In this case,
\begin{equation}
  \label{eq:extremal}
\min_{\Omega^{ab,\sigma}}  \sum_{a,b}
|\vvec^a - \vvec^b| \Omega^{ab,\sigma} 
= L(\Pcal^\sigma(h),\Ncal^\sigma(h)),
\end{equation}
where $\Pcal^\sigma(h)$ (resp.~$N^\sigma(h)$) contains vertices
$\vvec^a$ for which $w^{a\sigma}$ equals $+1$ (resp.~$-1$).  The case
of general nonzero wrapping numbers values is treated by including
$\vvec^a$ with multiplicity $|w^{a\sigma}|$ in either
$\Pcal^\sigma(h)$ (for $w^{a\sigma} > 0$) or $N^\sigma(h)$ (for
$w^{a\sigma} < 0$).  The same argument applies in every sector (there
is a separate
minimal connection for each $\sigma$), and (\ref{eq:E_-})
follows.

\end{proof}
\section{Upper bound for rectangular prisms}\label{sec: upper bound}

Let $P$ denote a rectangular prism centred at the origin of
three-dimensional Euclidean space.  We take the edges of $P$ to be
parallel to the coordinate axes and of lengths $L_x$, $L_y$, $L_z$,
oriented so that $ L_x \ge L_y \ge L_z$.  It will be convenient to
introduce the half-lengths
\begin{equation}
  \label{eq:half-length}
  l_j = L_j/2
\end{equation}
(here and in what follows, the index $j$ takes values $x$, $y$ or
$z$).
Then the vertices of $P$ are of the form
\begin{equation}
  \label{eq:vertices}
  \vvec^a = (\pm l_x, \pm l_y, \pm l_z),
\end{equation}
where the vertex label $a$ designates the signs in
(\ref{eq:vertices}).  



Let $O^a \subset S^2$ denote the spherical octant of
directions  about $\vvec^a$  which are contained in $P$.
Eg, for $\vvec^a = (-l_x,-l_y,-l_z)$, $O^a$ is the positive octant $\{
\svec \in S^2 \,|\, s_j \ge 0\}$.   
The boundary of $O^a$, $\partial O^a$, contains the directions 
which lie in the faces at $\vvec^a$, and 
is composed of quarter-segments of the great circles about  $\xhat$, $\yhat$
and $\zhat$.  Let $\partial O^a_j$ denote the segment about $\jhat$.

Choose $l$ so that $0< l \le
l_z$. Then  $\vvec^a + lO^a$ is contained in $P$, so that $\vvec^a + lO^a$
is a cleaved surface.  Given $\nvec \in \Ccal_P(h)$,
we can define a unit-vector field $\nuvec^a$ on $O^a$ by
\begin{equation}
  \label{eq:octant_config_example}
  \nuvec^a(\svec) = \nvec(\vvec^a + l\svec).
\end{equation}
Tangent boundary conditions imply that, for $\svec \in \partial
O^a_j$, $\nuvec^a(\svec)$ is orthogonal to $\jhat$.
Denote the set of $\nuvec^a$'s collectively by $\nuvec$.  The wrapping
numbers of $\nvec$, and hence its homotopy type, are determined by
$\nuvec$.
$\nuvec$ is an example of an {\it octant configuration}, which we define
generally as follows:
\begin{defn}
  An {\it octant configuration} $\nuvec$ with admissible topology $h =
  \{w^{a\sigma}\}$ is a set of continuous piecewise-smooth maps
  $\nuvec^a: O^a \rightarrow S^2$ satisfying tangent boundary
  conditions, 
  \begin{equation}
    \label{eq:nuvec_tangent_bc}
    \svec \in \partial O^a_j \implies \nuvec(\svec)\cdot \jhat = 0,
  \end{equation}
such that
\begin{equation}
      \label{eq:wrapping_for_nu}
      \int_{O^a} \nuvec^* (\chi^\sigma \omega) = w^{a\sigma} A^\sigma.
    \end{equation}
\end{defn}
The Dirichlet energy  of an octant configuration $\nuvec$ on the
octant $O^a$ is defined by
\begin{equation}
  \label{eq:E_2}
  E_{(2)}^a(\nuvec) = \int_{O^a} \left(\grad \nuvec^a
\right)^2(\svec) \, d\Omega^a.
\end{equation}
Here and in what follows, it will be convenient to regard $\grad
\nu_j^a(\svec)$ (the gradient of the $j$th component of $\nuvec^a$) 
as a vector in $\Rr^3$ which is tangent to $O^a$ at $\svec$.
$d\Omega^a$ in (\ref{eq:E_2}) denotes the area element on $O^a$ (normalised so that
$O^a$ has area $\pi/2$). 
The Dirichlet energy on the octant edge $\partial O^a_j$ is given by
\begin{equation}
  \label{eq:E_1^c}
  E_{(1)j}^{a}(\nuvec) = 
\int_0^{\pi/2} \left({\frac{d}{d\alpha}
    \nuvec^a}(\svec^a_j(\alpha))\right)^2 \,d\alpha.
\end{equation}
Here, $\svec^a_j(\alpha)$ denotes the
parameterisation of $\partial O^a_j$ by arclength (ie, angle)
$\alpha$.  For example, if $\vvec^a = (-l_x,-l_y,-l_z)$,
\begin{equation}
  \label{eq:O^a_j}
  \svec^a_j(\alpha) = \cos\alpha \khat + \sin\alpha \lhat,
  \ 0 \le \alpha \le \pi/2,
\end{equation}
where $(j,k,l)$ denote a triple of distinct indices.

By an {\it extension} of an octant configuration $\nuvec$, we mean a
continuous, piecewise-smooth unit-vector
field $\nvec$ on $P$ such that $ \nvec(\vvec^a + l \svec) = \nuvec^a(\svec)$
for all $\svec \in O^a$.  Obviously, if $\nuvec$ has topology $h$, so
has its extension $\nvec$. 
We introduce the following notation: Given functions $f$ and $g$ on a
domain $W$, we write $f \lesssim g$ to mean there exists a constant $C$
such that $|f| \le C |g|$ on $W$.  In this case, we say that $f$ is
dominated by $g$.
\begin{lem}\label{lem:lem.1}
  Let $\nuvec$ be an octant configuration with admissible
  topology $h$.  Then $\nuvec$ can be extended
  to a continuous piecewise-differentiable configuration $\nvec \in \Acal_P(h)$
  such that
    \begin{equation}
    \label{eq:prop lem.1}
    E(\nvec) \lesssim   \kappa^3 L_z \left( \sum_{a} E_{(2)}^a(\nuvec) 
  +     \sum_{aj} \left(E_{(1)j}^{a}(\nuvec)\right)^{1/2}\right).
  \end{equation}
\end{lem}
\noindent Theorem~\ref{thm:2} is proved by constructing octant configurations
whose Dirichlet energies on the octants and their edges scale appropriately
with the wrapping numbers.  These configurations are provided by the following:
\begin{lem}\label{lem:lem.2}
  Given an admissible topology $h = \{w^{a\sigma}\}$, there exists an
  octant configuration
$\nuvec$ with topology $h$ such that
  \begin{align}
    \sum_{a} E_{(2)}^a(\nuvec) &\lesssim \sum_{a\sigma} |w^{a\sigma}|,
    \label{eq:lem2_1}\\
  \sum_{aj} E_{(1)j}^{a}(\nuvec) &\lesssim \sum_{a\sigma} |w^{a\sigma}|^2.   \label{eq:lem2_2}
  \end{align}
\end{lem}
\noindent The proofs of Lemmas \ref{lem:lem.1} and \ref{lem:lem.2}, which
involve explicit constructions and estimates, are given in Sections
\ref{sec: pf of lemma 1} and \ref{sec: pf of lemma 2} respectively.

\begin{proof}[Proof of Theorem \ref{thm:2}]  Given an admissible topology $h =
  \{w^{a\sigma}\}$, we choose an octant configuration $\nuvec$ with
  topology $h$ as in Lemma~\ref{lem:lem.2}, and extend it to a
  unit-vector field $\nvec$ on $P$ as in Lemma~\ref{lem:lem.1}.
  From the Cauchy-Schwartz inequality, (\ref{eq:lem2_2}) implies that
  \begin{equation}
    \label{eq:E_1_square_root}
     \sum_{aj} (E_{(1)j}^{a}(\nuvec))^{1/2} \lesssim \sum_{a\sigma} |w^{a\sigma}|.
  \end{equation}
Together, (\ref{eq:prop  lem.1}), (\ref{eq:lem2_1}) and
(\ref{eq:E_1_square_root}) provide an estimate for $E(\nvec)$, and
therefore an upper bound for $E^\inff_P(h)$, 
\begin{equation}
  \label{eq:E(n)inproof}
  E^\inff_P(h) \le E(\nvec) \lesssim \kappa^3 L_z
  \sum_{a\sigma} |w^{a\sigma}|.
\end{equation}
From Theorem~\ref{thm:1}, a lower bound for $E^\inff_P(h)$ is given
by 
\begin{equation}
  \label{eq:prism_lower_bound}
  E^-_P(h) =  \sum_\sigma 2 \frac{\pi}{2} L(P^\sigma(h), N^\sigma(h))
\end{equation}
($A^\sigma = \pi/2$ for a rectangular prism).  The
minimum distance between vertices of $P$ is $L_z$.  As the number of elements of
$P^\sigma(h)$ (and of $N^\sigma(h)$) is $\half \sum_a |w^{a\sigma}|$,
it follows that
\begin{equation}
  \label{eq:bound_on_min_con}
   L(P^\sigma(h), N^\sigma(h)) \ge \half L_z \sum_{a} |w^{a\sigma}|.
\end{equation}
From (\ref{eq:E(n)inproof}) and (\ref{eq:bound_on_min_con}), we
conclude that
\begin{equation}
  \label{eq:final_statement_in_pf_of_1}
  E^\inff_P(h) \lesssim  \kappa^3 E^-_P(h).
\end{equation}
\end{proof}

We remark that the octant configurations of Lemma~\ref{lem:lem.2}
must be chosen with some care, as the following example illustrates (details may be
found in  \cite{majthesis}).  For simplicity, take $P$ to be the unit cube.  In the
(positive) octant about $\vvec^a = (-\half, -\half, -\half)$, let
\begin{equation}
  \label{eq:nuvec_example}
  \nuvec^a(\theta,\phi) = (\sin \alpha\cos\beta, \sin
  \alpha\sin\beta, \cos \alpha),
\end{equation}
where $ 0 \le \theta,\phi \le \pi/2$ and $\alpha = (4M+1)\theta$, $\beta = (4N+1)\phi$
for integers $M$ and $N$. Given $(x,y,z) \in P$ with $x,y,z \le
0$,  let $(\theta,\phi)$ denote the polar angles of $(x,y,z)$ with
respect to
$\vvec^a$, and let $\nvec(x,y,z) = \nuvec^a(\theta,\phi)$.  We define
$\nvec$ elsewhere via $\nvec(\pm x, y, z) = \nvec(x, \pm y, z) =
\nvec(x,y,\pm z) = \nvec(x,y,z)$ (so that $\nvec$ is a
reflection-symmetric configuration \cite{mrz2004b,mrz2006}).  Then 
$\nvec$ is continuous and  satisfies 
tangent boundary conditions.  Denote its homotopy class by $h_{MN}$.  It is straightforward to
compute the wrapping numbers (it turns out that they scale linearly with $M$ and $N$),
and, from Theorem~\ref{thm:1}, to obtain the following lower bound:
\begin{equation}
  \label{eq:lb_for_MN}
  E^-_P(h_{MN}) = (2\max(M+2N, 2M+N) +1)\frac{\pi}{4}.
\end{equation}

It turns out that $E(\nvec)$ can be evaluated exactly 
as a finite sum of Appell hypergeometric functions.  For large
$M$ and $N$, the energy is given asymptotically by
\begin{equation}
  \label{eq:E_n_asympt}
  E(\nvec) \sim 4\sqrt{3}((4M+1)^2 + \half \ln M (4N+1)^2\frac{\pi}{2}.
\end{equation}
Clearly $E(\nvec)$ 
does not scale linearly with $M$ and $N$, so is not dominated
by the lower bound $E^-_P(h_{MN})$.

\section{Extending octant configurations}\label{sec: pf of lemma 1}

\begin{figure}
\begin{center}
\includegraphics{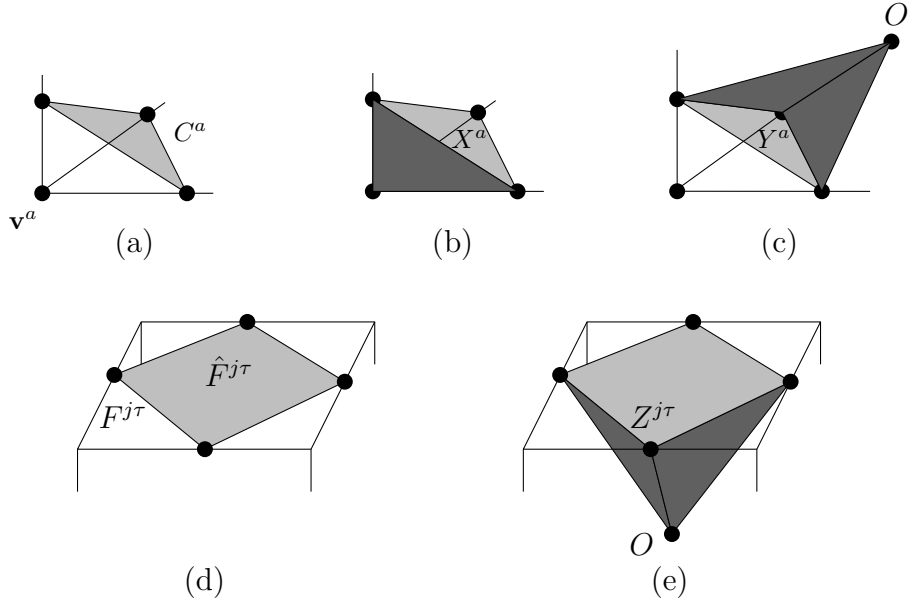}
\caption{(a) The cleaved plane $C^a$ (b) The pyramid $X^a$ (c) The
  pyramid $Y^a$ (d) The truncated face $F^{j\tau}$ (e) The pyramid $Z^{j\tau}$}
\label{fig: fig1}
\end{center}
\end{figure}

Let us specify the geometry of the prism $P$ in greater detail.  Let
\begin{equation}
  \label{eq:m^aj}
  \mvec^{a}_{(j)} = \vvec^a - v^a_j \jhat
\end{equation}
denote the midpoint of the edge through $\vvec^a$ along
$\jhat$ (here, $v^a_j$ is the $j$th component of $\vvec^a$).  Let $C^a$ denote the triangle whose vertices are the
midpoints of the edges coincident at $\vvec^a$,
\begin{equation}
  \label{eq:C^a}
  C^a = \{\rvec = \tau_x \mvec^{a}_{(x)} + \tau_y \mvec^{a}_{(y)} + \tau_z \mvec^{a}_{(z)} \,
  | \, \tau_j \ge 0, \sum_j \tau_j = 1\}.
\end{equation}
We call $C^a$ a {\it cleaved face}  (see Figure~\ref{fig: fig1}).
$\pvec \in C^a$ satisfies
\begin{equation}
  \label{eq:pvec}
  \Cvec^a\cdot\pvec = 2,
\end{equation}
where 
\begin{equation}
  \label{eq:Chat}
 \Cvec^a = \left( (v^a_x)^{-1}, (v^a_y)^{-1}, (v^a_z)^{-1}\right).
\end{equation}
is an (unnormalised) outward normal on $C^a$.
Let $h$ denote the distance from $C^a$ to the origin.  Then
\begin{equation}
  \label{eq:hdist}
\frac{2}{3} l_z \le  h = \frac{2}{|\Cvec^a|} < 2l_z.
\end{equation}
Let $F^{j\tau}$, where $\tau = \pm 1$, denote face of the prism which
lies in the 
plane $\{r_j = \tau l_j\}$.   
Let $\Fhat^{j\tau} \subset F^{j\tau}$ denote the rhombus whose 
vertices lie at the midpoints of the edges of $F^{j\tau}$. We call
$\Fhat^{j\tau}$ a {\it truncated face}  (see Figure \ref{fig: fig1}).



We partition $P$ into three sets of pyramids, denoted $X^a$, $Y^a$ and
$Z^{j\tau}$. 
$X^a$ and $Y^a$ have the cleaved face $C^a$ as their (shared) base.  $X^a$ has
its apex at $\vvec^a$, while $Y^a$ has its apex at the origin.
$Z^{j\tau}$ has the truncated face $\Fhat^{j\tau}$ as its base and its
apex at the origin.  Every point of $P$ belongs either to the interior
of just one of these pyramids or else to the boundary between two or
more of them  (see Figure~\ref{fig: fig1}). 

In the proof of Lemma~\ref{lem:lem.1}, we define $\nvec$, the extension of the octant
configuration $\nuvec$, to be constant along rays from 
$C^a$    to the origin 
(apart from a small neighbourhood
thereof) and along rays to $\vvec^a$.  We  then show that the energy of $\nvec$ in $X^a$ and $Y^a$
is proportional to $E^a_{(2)}(\nuvec)$.  The construction of $\nvec$ in
$Z^{j\tau}$ is more complicated.  On $\partial \Fhat^{j\tau}$ (the boundary of the base),
$\nvec$ is determined by $\nuvec$, and in the interior of
$\Fhat^{j\tau}$, we define $\nvec$ by a simple interpolation which respects
tangent boundary conditions.  In the interior of 
$Z^{j\tau}$, we do {\it not} take $\nvec$ to be constant along rays from the apex
to the base, as this would give rise to an energy proportional
to $\sum_{\vvec^a \in F^{j\tau}} E_{(1)j}^a(\nuvec)$, which, for the
octant configurations of Lemma~\ref{lem:lem.2}, would scale as the
square of the wrapping numbers.  Instead, along such rays, and over a distance
\begin{equation}
  \label{eq:sigma}
  \sigma= \left(\pi \sum_{\vvec^a\in F^{j\tau}} (E_{(1)j}^{a}(\nuvec))\right)^{-1/2},
\end{equation}
$\nvec$ is rotated toward the normal $\jhat$.  This leads to an energy
in $Z^{j\tau}$ proportional to $1/\sigma$.

\begin{proof}[Proof of Lemma \ref{lem:lem.1}.]
Given an octant configuration $\nuvec$, we define an extension $\nvec$
 on the pyramids 
 $X^a$, $Y^a$ and $Z^{j\tau}$ (Steps 1 -- 3) with Dirichlet energies $E_{X^a}(\nvec)$,
 $E_{Y^a}(\nvec)$ and $E_{Z^{j\tau}}(\nvec)$ bounded as follows:
\begin{subequations} \label{eq:prism_bounds}
\begin{align}
    E_{X^a}(\nvec)&\le  \half \kappa L_z
    E_{(2)}^a(\nuvec),\label{eq:X bound}\\ 
 E_{Y^a}(\nvec)&\lesssim \kappa^3 L_z
    E_{(2)}^a(\nuvec),\label{eq:Y bound}\\ 
   E_{Z^{j\tau}}(\nvec)&\lesssim  \kappa^2 L_z
    \sum_{\vvec^a\in F^{j\tau}} (E_{(1)j}^{a}(\nuvec))^{1/2},\label{eq:Z bound}
\end{align}
\end{subequations}
Then
\begin{multline}
  \label{eq:sum_of_bounds}
  E(\nvec) = \sum_a \left(E_{X^a}(\nvec) +  E_{Y^a}(\nvec)\right) + \sum_{j\tau}
  E_{Z^{j\tau}}(\nvec)\\
\lesssim \kappa^3 L_z \left(\sum_a  E_{(2)}^a(\nuvec) + \sum_{aj}  \left(E_{(1)j}^{a}(\nuvec)\right)^{1/2}\right).
\end{multline}
To ensure continuity (Step 4) we modify the construction of $\nvec$ near the
origin while preserving the bounds (\ref{eq:prism_bounds}).\\

\noindent {\it Step 1.  Construction in $X^a$}\quad   From (\ref{eq:C^a}) and
(\ref{eq:pvec}), 
points in $X^a$ are of the form
$\vvec^a +  r\svec$, where $\svec \in O^a$ and 
$0 < r \le r^a(\svec)$ with 
 \begin{equation}
   \label{eq:r_max}
   r^a(\svec) = \frac{1}{|\Cvec^a\cdot\svec|};
 \end{equation}
$r^a(\svec)$ is the distance from $\vvec^a$ to $C^a$ along $\svec$.
The maximal distance is half the length of the longest edge, so that
\begin{equation}
  \label{eq:r_max bound}
  r^a(\svec) \le r^a(\xhat) = l_x. 
\end{equation}
We define $\nvec$ in $X^a$ by
 \begin{equation}
   \label{eq:nvec_in_X_vec}
   \nvec(\vvec^a + r \svec) = \nuvec^a(\svec), \quad \svec \in O^a,\ 0 < r \le r^a(\svec).
 \end{equation}
Then, from (\ref{eq:E_2}) and (\ref{eq:r_max bound}),
\begin{multline}
  \label{eq:E_X^a}
  E_{X^a}(\nvec) := \int_{X^a} (\grad \nvec)^2 \, dV = \int_{O^a}
  r^a(\svec) (\grad \nuvec^a(\svec))^2 \,d\Omega^a\\
  \le l_x E_{(2)}^a(\nuvec) \le \half \kappa L_z E_{(2)}^a(\nuvec),
\end{multline}
as in (\ref{eq:X bound}).\\


\noindent{\it Step 2. Construction in $Y^a$}\quad Points in $Y^a$ (excluding the
origin) are of the form
$\lambda \pvec$, where $\pvec \in C^a$ and $ 0 < \lambda \le 1$.  
We define $\nvec$ in $Y^a$ by
\begin{equation}
  \label{eq:n in Y^a}
  \nvec(\lambda \pvec) = \nvec(\pvec).
\end{equation}
Note that $\nvec(\pvec)$ is fixed in Step 1, since $C^a$
belongs to $X^a$ as well as $Y^a$.  

To estimate
$E_{Y^a}(\nvec)$, we resolve $\grad \nvec$ into components
tangent and normal to the cleaved face $C^a$,
\begin{equation}
  \label{eq:grad_X_1}
  \grad 
\nvec = \gradt \nvec + \gradn \nvec,
\end{equation}
so that  $(\Cvec^a\cdot\gradt)\nvec = 0$ and $(\tvec \cdot \gradn)
\nvec = 0$ for $\tvec$ tangent to $C^a$.
From (\ref{eq:n in Y^a}), $\gradt \nvec(\lambda \pvec) = \lambda^{-1}
\gradt \nvec(\pvec)$.  Therefore,
\begin{equation}
  \label{eq:gradt estimate}
  (\gradt \nvec(\lambda \pvec))^2 = \frac{1}{\lambda^2} (\gradt \nvec(\pvec))^2.
\end{equation}
To estimate $(\gradn \nvec)^2$, we note that $ (\pvec \cdot
\grad) \nvec (\lambda \pvec) = 0$ ($\nvec$ is constant along rays
in $Y^a$ through the origin)
and resolve $\pvec$ into components $\pvect$ and $\pvecn$ tangent
and normal to $C^a$ to obtain
\begin{equation}
  \label{eq:gradn_n}
  (\gradn \nvec(\lambda \pvec))^2 \le 
   \frac{|\pvect|^2}{|\pvecn|^2} (\gradt\nvec(\lambda \pvec))^2 \le
   \frac{|\pvect|^2}{|\pvecn|^2} \frac{1}{\lambda^2} (\gradt\nvec( \pvec))^2. 
\end{equation}
Since $(\gradt\nvec( \pvec))^2 \le (\grad\nvec( \pvec))^2$,
(\ref{eq:gradt estimate}) and
(\ref{eq:gradn_n}) together give 
\begin{equation}
  \label{eq: grad n est in Y temp}
  (\grad \nvec(\lambda\pvec))^2 = 
  \frac{|\pvec|^2}{|\pvecn|^2}  \frac{1}{\lambda^2} (\grad
  \nvec(\pvec))^2.
\end{equation}
Clearly $|\pvec| \le l_x$ while $|\pvecn|$ is just $h$, the distance 
from $C^a$ to the origin, and $h > 2l_z/3$ (cf (\ref{eq:hdist})), so that
\begin{equation}
  \label{eq: grad n est in Y temp2}
  (\grad \nvec(\lambda\pvec))^2 
\le \frac{9}{4}\kappa^2
 \frac{1}{\lambda^2} (\grad
  \nvec(\pvec))^2.
\end{equation}

The volume element on $Y^a$ is given by
\begin{equation}
  \label{eq:X_volume}
  dV = h{\lambda^2}\, d^2 c d \lambda,
\end{equation}
where $d^2 c$ is the Euclidean area element on $C^a$.
Since $h < 2l_z$, 
\begin{equation}
  \label{eq:energy_X_calc}
E_{Y^a}(\nvec) =  \int_{Y^a} (\grad \nvec)^2\, dV 
< \frac{9}{2} \kappa^2 l_z\int_{C^a} (\grad \nvec)^2(\pvec)\, d^2
c.
\end{equation}
Letting $\svec = (\pvec - \vvec)/|\pvec - \vvec|$, we can write the
preceding as an integral over $O^a$. We have that
\begin{equation}\label{eq: P->O^a}
  d^2 c = \frac{|\pvec|^2}{|\svec \cdot \Cvec^a|/|\Cvec^a|}d\Omega^a,\quad 
(\grad \nvec(\pvec))^2 = \frac{1}{|\pvec|^2} (\grad \nuvec^a(\svec))^2,
\end{equation}
and
\begin{equation}
  \label{eq:cos_inequality}
  \frac{|\Cvec^a|}{|\svec\cdot\Cvec^a|} < \frac{3/l_z}{1/l_x} = 3\kappa,
\end{equation}
so that (\ref{eq:energy_X_calc}) becomes
\begin{equation}
  \label{eq:energy_X_calc_2}
E_{Y^a}(\nvec) 
\le \frac{27}{2} \kappa^3 l_z \int_{O^a} (\grad \nuvec^a(\svec))^2\,
d\Omega^a \lesssim \kappa^3 L_z E^a_2(\nuvec),
\end{equation}
as in (\ref{eq:Y bound}).\\

\noindent
{\it Step 3. Construction in $Z^{j\tau}$} \quad 
To simplify the discussion and the notation, let us fix our attention
on the top face of the prism, with $j = z$ and $\tau = 1$, and
henceforth drop the designation $j\tau$, writing $Z$ for $Z^{j\tau}$,
$\Fhat$ for $\Fhat^{j\tau}$, etc, in what follows (the other faces are
handled similarly).  $\partial \Fhat$ may be parameterised as
$\Rvec(\phi) = (R(\phi)\cos\phi,R(\phi)\sin\phi, l_z)$, where
\begin{equation}
   \label{eq:R(alpha)}
   R(\phi) = \frac{l_x l_y}{l_y |\cos\phi| + l_x |\sin\phi|}
 \end{equation}
 and $\phi$ is the polar angle about the $z$-axis.  On $\partial
 \Fhat$ (which is also contained in the cleaved faces), $\nvec$ is
 defined in Step 1.
It follows that $\nvec$ is continuous  on $\partial
 \Fhat$ (including the midpoints of the edges
 of $F$, which belong to two cleaved faces, as $\nuvec$ has an
 admissible topology) and satisfies 
tangent boundary conditions there.  Tangent boundary conditions imply that
\begin{equation}
  \label{eq:Theta}
  \nvec(\Rvec(\phi)) =
\epsilon \cos(\Theta(\phi)) \yhat + \sin(\Theta(\phi)) \xhat,
\end{equation}
where $\Theta(\phi)$ may be taken to be continuous and piecewise
smooth, with $\Theta(2\pi) - \Theta(0)$ equal to a multiple of $2\pi$.  
Since $\nuvec$ has an admissible topology, we must have $\Theta(2\pi)
= \Theta(0)$.
$\epsilon = \pm 1$ can be chosen so that 
\begin{equation}
  \label{eq:Theta(0)}
  \Theta(0) = \Theta(2\pi) = 0.
\end{equation}

We introduce 
polygonal cylindrical coordinates
$(h,\xi,\phi)$  on $Z$, defined by
\begin{gather}
  x = h \xi R(\phi) \cos\phi, \quad y = h\xi R(\phi) \sin
  \phi, \quad z = l_z h, \nonumber \\ 
0 < h \le 1, \quad 0 \le \xi \le 1, \quad 0 \le \phi < 2\pi.  \label{eq:Z coords}
\end{gather}
$\xi$, the radial coordinate, is scaled to equal $1$ on the sides of
$Z$ and $0$ along the $z$-axis.  
Then let
\begin{equation}
  \label{eq:N_(2)}
  \Nvec_{(2)}(\xi,\phi) = \epsilon \cos(\xi\Theta) \yhat + \sin(\xi\Theta)\xhat.
\end{equation}
$ \Nvec_{(2)}$ gives a continuous extension of $\nvec$ to the interior of
$\Fhat$ which satisfies tangent boundary conditions.  
A continuous extension to the interior of $Z$ 
is given by
\begin{equation}
  \label{eq:N}
  \Nvec(h,\xi,\phi) = \cos \gamma  \Nvec_{(2)} + \sin\gamma \zhat,
\end{equation}
where $\gamma = \gamma(h,\xi)$ is given by
\begin{equation}
  \label{eq:beta}
  \gamma(h,\xi) = \Phi\left(\frac{1-h}{\sigma}\right)
\Phi\left(\frac{1-\xi}{\sigma}\right)\frac{\pi}{2}, \quad \Phi(s) = 
  \begin{cases}
    s, & s < 1,\\
    1, & s \ge 1,
  \end{cases}
\end{equation}
and $0 < \sigma < 1$. 
Thus, $\gamma$ vanishes on the boundary
of $Z$ and has a constant value, $\pi/2$, at interior points
sufficiently far from the boundary. $\sigma$, which determines how far, will be specified below.
We define $\nvec$ on $Z$ as 
\begin{equation}
  \label{eq:N def}
  \nvec(x(h,\xi,\phi),y(h,\xi,\phi),z(h,\xi,\phi)) = \Nvec(h,\xi,\phi).
\end{equation}
It is readily checked that (\ref{eq:N def}) agrees with (\ref{eq:n in
  Y^a}) at points on the boundary of $Z$ except at the origin.

The energy of $\nvec$ in $Z$ is given by
 $E_{Z}(\nvec)$ as follows.  
In terms of the coordinates $(h,\xi,\phi)$, we have that
 \begin{multline}
   \label{eq:E_Z}
   E_{Z}(\nvec) = \int_{Z} (\grad \nvec)^2 \,dV\\
 = \int_0^1 dh \int_0^1 d\xi \int_0^{2\pi} d\phi \,
 \big|
 (\grad \xi \times \grad\phi)\cdot \grad h
 \big|^{-1} \left(
 \Nvec_h \grad h + \Nvec_\xi \grad \xi + \Nvec_\phi \grad \phi
 \right)^2,
 \end{multline}
where $\Nvec_h = \partial \Nvec/\partial h$, $\Nvec_\xi = \partial
\Nvec/\partial \xi$, 
etc.  From 
(\ref{eq:N_(2)}) and (\ref{eq:N}), we get that
\begin{align}\label{eq: partials of Nvec}
\Nvec_h &= \gamma_h (\cos\gamma \zhat - \sin\gamma
\Nvec_{(2)}),\nonumber\\
\Nvec_\xi &= \gamma_\xi (\cos\gamma \zhat - \sin\gamma \Nvec_{(2)})
+ \epsilon\Theta \cos\gamma  \Nvec_{(2)}\times \zhat ,\nonumber\\
\Nvec_\phi &= 
\epsilon\xi \Theta' \cos\gamma  \Nvec_{(2)}\times \zhat.
\end{align}
From (\ref{eq:Z coords}), we get that
\begin{align}\label{eq: grad of coordinates}
  \grad h &= (0,0,1/l_z),\nonumber\\
  \grad \xi &= \xi ((R\cos\phi + {R}'\sin\phi)/(\rho R),
(R\sin\phi - {R}'\cos\phi)/(\rho R), -1/(l_z h)),\nonumber\\
\grad\phi &= (-\sin\phi,\cos\phi,0)/\rho,
\end{align}
where $\rho = (x^2 + y^2)^{1/2} = h\xi R$.  Straightforward calculation then gives an expression for
$E_{Z}(\nvec)$ of the form
\begin{equation}
  \label{eq:E_Z2}
  E_{Z}(\nvec) = \sum_{i=1}^5 \Ecal_i = \sum_{i=1}^5
\int_0^1 dh \int_0^1 d\xi \int_0^{2\pi} d\phi \, I_i,
\end{equation}
where the integrands for the separate contributions $\Ecal_i$ are given by 
\begin{gather}\label{eq: I's}
  I_1 = l_z \cos^2\gamma \xi \Theta'^2,\quad 
  I_2 = \frac{h^2}{l_z} \gamma_h^2 \xi {R}^2,\quad
  I_3 = -2\frac{h}{l_z} \gamma_h \gamma_\xi \xi^2 {R}^2,\nonumber\\
  I_4 = -2\epsilon l_z \cos^2\gamma \xi \frac{{R}'}{R} \Theta \Theta',\nonumber\\
  I_5 = \left(l_z \xi \left( 1 + \frac{{R}'^2}{{R}^2}\right) +
  \frac{\xi^3}{l_z} {R}^2\right)
\left(
\cos^2\gamma \Theta^2 + \gamma_\xi^2
\right).
\end{gather}
We consider these contributions in turn.

Concerning $\Ecal_1$, since
$\cos^2\gamma$ vanishes for $0 < \xi, h < 1 - \sigma$ (cf (\ref{eq:beta})),
it follows that
\begin{equation}
  \label{eq:E_1-3}
   \Ecal_{1} \le 2 l_z \sigma \int_0^{2\pi} \Theta'^2(\phi) \, d\phi. 
\end{equation}
The integral of ${\Theta'}^2(\phi)$ can be related to the Dirichlet
edge energies $E_{(1)z}^a(\nuvec)$ for the vertices $\vvec^a$ which
lie on $F$.  For convenience, label these anticlockwise by $a =
0,1,2,3$ so that
\begin{equation}
  \label{eq:N_and_n}
  \nvec(\Rvec(\phi)) = \nuvec^a(\svec_z^a(\alpha(\phi)), \ (a-1)\pi/2 <
  \phi < a\pi/2,
\end{equation}
where $\svec_z^a(\alpha)$ parameterises the octant edge $\partial
O^a_z$ as in (\ref{eq:O^a_j}), and
$\alpha(\phi)$ is given 
by 
\begin{equation}
  \label{eq:phi and alpha}
  \tan \alpha = 
\begin{cases}
(l_x/l_y)^2 \tan\phi,& 0 \le \phi < \pi/2\ \text{or}\  \pi \le \phi < 3\pi/2, \\
 (l_y/l_x)^2\tan\phi,& \pi/2 \le \phi < \pi \ \text{or}\  3\pi/2 \le \phi < 2\pi.
\end{cases}
\end{equation}
($\phi$ is the angle with respect to the centre of $F$ and $\alpha$,
with $0 <  \alpha < \pi/2$,
the angle with respect consecutive vertices. (\ref{eq:phi and alpha})
gives the elementary relation between them.  See also Figure~\ref{fig:
  fig2}.)  Recalling (\ref{eq:Theta}), 
we get that
\begin{equation}
  \label{eq:Theta'^2}
  \Theta'^2(\phi) = \left(\frac{d}{d\phi} \nvec(\Rvec(\phi))\right)^2
  = \left.\left(\frac{d}{d\alpha}
  \nuvec^a(\svec_z^a(\alpha))\right)^2\right|_{\alpha = \alpha(\phi)}
  \left(
\frac{d\alpha}{d\phi}\right)^2 .
\end{equation}

\begin{figure}
\begin{center}
\includegraphics{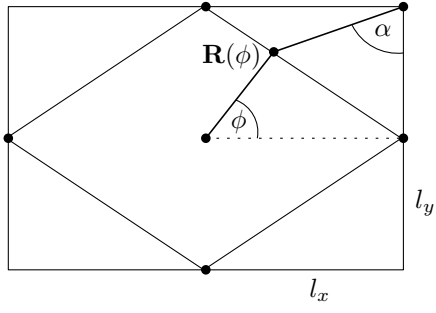}
\caption{The angles $\alpha$ and $\phi$}
\label{fig: fig2}
\end{center}
\end{figure}
It follows that
\begin{equation}
  \label{eq:int theta'}
  \int_0^{2\pi} {\Theta'}^2(\phi)\,d\phi = \sum_{\vvec^a \in F} 
  \int_0^{\pi/2}  \left(\frac{d}{d\alpha}
  \nuvec^a(\svec_z^a(\alpha))\right)^2
\left
  (\frac{d\phi}{d\alpha}\right)^{-1} \,d\alpha.
\end{equation}
From (\ref{eq:phi and alpha}) one has that
\begin{equation}
  \label{eq:bound on dphi/dalpha}
  \left| \left(\frac{d\phi}{d\alpha}\right)^{-1}\right| \le
  \frac{l_x^2}{l_y^2} \le \kappa^2 \
\end{equation}
Therefore,
\begin{equation}
  \label{eq:int theta'2}
  \int_0^{2\pi} {\Theta'}^2(\phi)\,d\phi \le \kappa^2 \sum_{\vvec^a    \in F}
E_{(1)z}^{a}(\nuvec).
\end{equation}
Substituting into (\ref{eq:E_1-3}), we get
\begin{equation}
  \label{eq:E_1-4}
   \Ecal_{1} \lesssim\kappa^2  l_z \sigma 
\sum_{\vvec^a    \in F}
E_{(1)z}^{a}(\nuvec).
\end{equation}

Next we consider 
\begin{equation}
  \label{eq:Ecal_2}
  \Ecal_{2} = \int_0^1 dh \frac{h^2}{l_z} \int_0^1 d\xi\,   
\xi \gamma_h^2\int_0^{2\pi} d\phi \, {R}^2.
\end{equation}
From (\ref{eq:R(alpha)}), $R(\phi) \le l_x$, and from (\ref{eq:beta}), 
$|\gamma_h| $ vanishes for $h < 1-\sigma$ and is bounded by
$\pi/(2\sigma)$ for $h > 1- \sigma$. 
Therefore,
\begin{equation}
  \label{eq:E_2-2}
   \Ecal_{2}  < \frac{\pi^3}{4} \frac{l_x^2}{l_z} \frac{1}{\sigma} \lesssim 
\kappa^2 l_z \frac{1}{\sigma}.
\end{equation}
We estimate $\Ecal_3$ similarly;  noting that $|\gamma_\xi|$ vanishes
for  $\xi < 1-\sigma$ and is bounded by
$\pi/(2\sigma)$ for $\xi > 1- \sigma$, we obtain
\begin{equation}
  \label{eq:E_3Z}
  |\Ecal_3| \lesssim   \kappa^2 l_z\frac{1}{\sigma}.
\end{equation}

$\Ecal_4$ is given by
\begin{equation}
  \label{eq:E_Z4}
   \Ecal_4 = -2\epsilon l_z\int_0^1 dh \int_0^1 d\xi\, \xi \cos^2\gamma 
\int_0^{2\pi} d\phi \, \Theta \Theta' \frac{{R}'}{R}.
\end{equation}
We consider the $\phi$-integral first.  From the Cauchy-Schwartz
inequality,
\begin{equation}
  \label{eq:Theta Theta'2}
  \left |\int_0^{2\pi} \Theta \Theta' \frac{{R}'}{R} \, d\phi \right |\le
\left(
\int_0^{2\pi}
\Theta^2\,d\phi
\right)^{1/2}
\left(
\int_0^{2\pi}
\Theta'^2 \left(\frac{R'}{R}\right)^2\,d\phi
\right)^{1/2}.
\end{equation}
In the first factor on the right-hand side, note that,
from (\ref{eq:Theta}) and (\ref{eq:N_and_n}),
\begin{equation}
  \label{eq:Theta^2(alpha)}
  |\Theta(\phi)| \le 
\sum_{\vvec^a \in F}
\int_0^{\pi/2}
  \left|\left(\frac{d}{d\alpha}
  \nuvec^a(\svec_z^a(\alpha))\right)\right|
\,d\alpha, \quad  \text{for $0 \le \phi \le 2\pi$}.
\end{equation}
The Cauchy-Schwartz inequality then implies that
\begin{equation}
  \label{eq:Theta^2(alpha)2}
\Theta^2(\phi)  \le 2\pi 
\sum_{\vvec^a\in F} 
E_{(1)z}^{a}(\nuvec), \quad  \text{for $0 \le \phi \le 2\pi$}.
\end{equation}
In the second factor on the right-hand side of (\ref{eq:Theta
  Theta'2}), we have that $|R'/R | \le \kappa$ (cf
(\ref{eq:R(alpha)})), so it follows from (\ref{eq:int theta'2}) that
\begin{equation}
  \label{eq:second_integral_in_E4}
  \int_0^{2\pi}
\Theta'^2 \left(\frac{R'}{R}\right)^2\,d\phi
 \le \kappa^4 \sum_{\vvec^a    \in F}
E_{(1)z}^{a}(\nuvec).
\end{equation}
Substituting (\ref{eq:Theta^2(alpha)2}) and
(\ref{eq:second_integral_in_E4}) into (\ref{eq:Theta Theta'2}), we get
that
\begin{equation}
  \label{eq:phi-integral-2}
   \left|\int_0^{2\pi} \Theta \Theta' \frac{{R}'}{R} \, d\phi\right|
   \lesssim  \kappa^2\sum_{\vvec^a\in F} 
E_{(1)z}^{a}(\nuvec).
\end{equation}
We substitute (\ref{eq:phi-integral-2}) into (\ref{eq:E_Z4}) and
recall 
that $\cos^2\gamma$ vanishes for $0
< \xi, h < 1 - \sigma$ to get that
\begin{equation}
  \label{eq:EZ4-2}
  |\Ecal_{4}|  \lesssim
\kappa^2  l_z \sigma
\sum_{\vvec^a\in F} 
E_{(1)z}^{a}(\nuvec).
\end{equation}

Finally, we consider
\begin{equation}
  \label{eq:E_Z5}
   \Ecal_{5} = \int_0^1 dh \int_0^1 d\xi \int_0^{2\pi} d\phi\,
\left(
\cos^2\gamma \Theta^2 + \gamma_\xi^2
\right)
\left(l_z \xi \left( 1 + \frac{{R}'^2}{{R}^2}\right) +
  \frac{\xi^3}{l_z} {R}^2\right).
\end{equation}
Let us first estimate the $\phi$-dependent terms.  
From (\ref{eq:R(alpha)}), we have that $R \le \kappa l_z$ and  $|R'/R| \le \kappa$, while
(\ref{eq:Theta^2(alpha)2}) provides a bound for $\Theta^2(\phi)$.
Substituting into (\ref{eq:E_Z5}), we get that
\begin{equation}
  \label{eq:E_Z5-2}
  \Ecal_{5} \le  2\pi l_z \int_0^1 dh \int_0^1 d\xi 
\left( 
2\pi \cos^2 \gamma \sum_{\vvec^a\in F} 
E_{(1)z}^{a}(\nuvec) + \gamma_\xi^2
\right)
\left(
 \xi(1 + \kappa^2) + \xi^3 \kappa^2)
\right).
\end{equation}
Recalling that $\cos^2\gamma$ and $\gamma_\xi^2$ vanish for $0 < \xi,
h < 1 - \sigma$ while $\gamma_\xi^2$ is bounded by $\pi/(2\sigma)$, 
we obtain the bound
\begin{multline}\label{eq:E_Z5-3}
  \Ecal_{5} \le 4\pi l_z \sigma 
 \left(2\pi 
\sum_{\vvec^a\in F} 
E_{(1)z}^{a}(\nuvec) + \frac{\pi^2}{4\sigma^2} \right) 
(1 + 2\kappa^2)\\ \lesssim
\kappa^2 l_z \sigma \left(
\sum_{\vvec^a\in F} 
E_{(1)z}^{a}(\nuvec)\right) + \kappa^2 l_z \frac{1}{\sigma}.
\end{multline}

We substitute the estimates (\ref{eq:E_1-3}), (\ref{eq:E_2-2}),
(\ref{eq:E_3Z}), (\ref{eq:EZ4-2}) and (\ref{eq:E_Z5-2}) for $\Ecal_i$ into
(\ref{eq:E_Z2}), and take
\begin{equation}
  \label{eq:sigma scaling}
  \sigma = \left(\pi \sum_{\vvec^a\in F} 
E_{(1)z}^{a}(\nuvec)\right)^{-1/2}.
\end{equation} 
We can verify that $\sigma < 1$ as follows:
Tangent boundary conditions imply that, 
for $\phi$ a multiple of $\pi/2$ (ie, for $\Rvec(\phi)$ belonging to an edge
of $F$),
$\Theta(\phi) = \pi/2 \mod
2\pi$.  Then   (\ref{eq:Theta^2(alpha)2}) implies that 
$\sum_{\vvec^a  \in F}
E_{(1)z}^a \ge \pi/8$, so that $\sigma < \sqrt{8}/\pi < 1$.
Then (\ref{eq:E_Z5-3}) and (\ref{eq:sigma scaling}) give
\begin{equation}\label{eq: E_Z f2}
 E_{Z} \lesssim \kappa^2 l_z
\left(\sum_ {\vvec^a \in F} 
E_{(1z)}^{a}\right)^{1/2} \lesssim 
 \kappa^2 l_z 
\sum_{\vvec^a \in F} 
(E_{(1z)}^{a})^{1/2},
\end{equation}
as in (\ref{eq:Z bound}).
The same estimate may be carried out for the other pyramids
$Z^{j\tau}$.   Different lengths $l_j$ appear as appropriate, 
but since ratios of lengths are bounded by
$\kappa$, (\ref{eq: E_Z f2}) holds generally.\\

\noindent{\it Step 4. Continuity}\quad 
As defined, $\nvec$ is continuous everywhere except at the origin.
Here we modify 
$\nvec$ in a small neighbourhood of the origin to remove the
discontinuity while preserving the estimate (\ref{eq:prism_bounds}).

From the definitions (\ref{eq:n in Y^a}) and (\ref{eq:N def}) of
$\nvec$ in $Y^a$ and $Z^{j\tau}$,
$\nvec(\rvec)$ is radially constant in the ball $B_\delta$ about the
origin of radius $\delta =
(1-\sigma)l_z$. 
Let $\gammavec:S^2 \rightarrow S^2$ denote the restriction of
$\nvec$ to $\partial B_\delta$, given by
\begin{equation}
  \label{eq:gamma_defn}
  \gammavec(\svec) = \nvec(\delta \svec), \quad \svec \in B_\delta.
\end{equation}
Let
\begin{equation}
  \label{eq:E_gamma}
  E_{(2)}(\gammavec) = \int_{S^2} (\grad \gammavec)^2 d\Omega
\end{equation}
denote the Dirichlet energy of $\gammavec$.  Then, for $0 < \epsilon <
1$, the energy of $\nvec$ in the $\epsilon\delta$-ball about the
origin is given by
\begin{equation}
  \label{eq:B_rho_energy}
  E_{B_{\epsilon \delta}}(\nvec) = \epsilon \delta  E_{(2)}(\gammavec).
\end{equation}

$\gammavec$ is piecewise smooth, and, since
$\nuvec$ has an admissible topology, of degree zero.  It follows that
$\gammavec$ is smoothly homotopic to a constant map. Let $\Gammavec_s$ denote a
homotopy, so that  $\Gammavec_1(\svec) =
\gammavec(\svec)$ and $\Gammavec_0(\svec) = \svec_0$ . Let $\gvec$ be the unit-vector field on 
$B_\delta$
given by
\begin{equation}
  \label{eq:Gvec}
  \gvec(r\svec) = \Gammavec_{r/\delta}(\svec), 
\end{equation}
and let $ E_{B_\delta}(\gvec) $ denote its Dirichlet energy.  For $0 <
\epsilon < 1$, we define $\gvec_\epsilon$ to be the unit-vector field
on $B_{\epsilon \delta}$ given by 
\begin{equation}
  \label{eq:g_epsilon}
  \gvec_\epsilon(\rvec) = \gvec(\rvec/\epsilon).
\end{equation}
Then the energy of $\gvec_\epsilon$ in the $\epsilon\delta$-ball about the
origin is given by
\begin{equation}
  \label{eq:E_g_eps}
  E_{B_{\epsilon\delta}}(\gvec_\epsilon) = \epsilon  E_{B_\delta}(\gvec). 
\end{equation}

For any $0 < \epsilon < 1$ we can redefine $\nvec$ on 
$B_{\epsilon\delta}$, taking it to be $\gvec_\epsilon$ there, and
leaving $\nvec$  unchanged elsewhere.  It is clear 
that $\nvec$ as redefined is continuous and piecewise smooth. From
(\ref{eq:B_rho_energy}) and (\ref{eq:E_g_eps}), the redefinition
changes its energy by 
$\epsilon ( E_{B_\delta}(\gvec) -  \delta
E_{(2)}(\gammavec))$. As $\epsilon$ can be made arbitrarily small, the
estimates (\ref{eq:prism_bounds}) remain valid.

\end{proof}

\section{Constructing octant configurations}\label{sec: pf of lemma 2}
For definiteness, we consider the configuration on 
a particular octant, namely the positive octant about 
the vertex $(-l_x,-l_y,-l_y)$;  the treatment 
for the other octants is analogous.  To simplify the notation, we
will drop the vertex label $a$.  Hence, throughout this section, 
we write  $O =
\{\svec \in S^2 | s_j \ge 0\}$ for the positive octant (instead of
$O^a$),
$\nuvec: O \rightarrow S^2$  for the configuration on $O$ 
(instead of $\nuvec^a$),
$w^\sigma$ instead of $w^{a\sigma}$,  etc.
With
reference to (\ref{eq:E_2}) and (\ref{eq:E_1^c}), we let 
\begin{align}
    E_{(2)}(\nuvec) &= \int_{O} \left(\grad \nuvec
\right)^2(\svec) \, d\Omega,\label{eq: E_2 sans a}\\
  E_{(1)j}(\nuvec) &=
\int_0^{\pi/2} \left(\frac{d\nuvec}{d\alpha}\right)^2(\svec_j(\alpha)) \,d\alpha,\label{eq: E_1 sans a}
\end{align}
where $d\Omega$ denotes the area element on $O$ and 
\begin{equation}
  \label{eq:evec_j}
  \svec_j(\alpha) = \cos \alpha \khat + \sin \alpha \lhat
\end{equation}
denotes the parameterisation of $\partial O_j$.
Lemma~\ref{lem:lem.2} follows from showing that
  \begin{align}
    E_{(2)}(\nuvec) &\lesssim \sum_\sigma |w^\sigma|,
    \label{eq:lem2_1 sans a}\\
  \sum_j E_{(1)j}(\nuvec) &\lesssim \sum_{\sigma} |w^{\sigma}|^2,
  \label{eq:lem2_2 sans a}
  \end{align}
as analogous relations hold for the other
octants.  Before establishing (\ref{eq:lem2_1 sans a}) and (\ref{eq:lem2_2
  sans a}) in Section~\ref{sec: pf of lem2}, we first review the
topological characterisation of octant configurations (Section \ref{sec: top})
and their representation by complex functions, particularly conformal representatives
(Section \ref{sec: con}).

\subsection{Topological characterisation}\label{sec: top}

As discussed in \cite{rz2003} (in the context of general convex
polyhedra) and in \cite{mrz2004b, mrz2006} (for a rectangular prism),
the homotopy class of $\nuvec: O \rightarrow S^2$ may be characterised by certain
invariants, namely the {\it edge signs}, denoted $e = (e_x,e_y,e_z)$,
{\it kink numbers}, denoted $k = (k_x,k_y,k_z)$ and {\it trapped
  area}, denoted $\Omega$.  Here we recall the definitions of these
invariants and some relevant results for prisms; details may be found in the
references.

Tangent boundary conditions imply that $\nuvec(\jhat)$
is parallel to $\jhat$; the edge sign $e_j$ determines their relative
sign, ie
\begin{equation}
  \label{eq:edge sign}
  \nuvec(\jhat) = e_j \jhat.
\end{equation}
Tangent boundary conditions also imply that along $\svec_j(\alpha)$,
$\nuvec$ takes values in the $(kl)$-plane.  The integer-valued kink number
$k_j$ counts the number of windings of $\nuvec$ in this plane relative
to the minimum possible winding (a net rotation of $\pm \pi/2$), for
which $k_j = 0$.  
The trapped area $\Omega$ is the oriented area of the image of $\nuvec$, given
by
\begin{equation}
  \label{eq:trapped_area_1}
  \Omega = \int_{O} \nuvec^*\omega.
\end{equation}

For a rectangular prism, the sectors are octants of $S^2$ labeled by a
triple of signs $\sigma = (\sigma_x,\sigma_y,\sigma_z)$, with
$S^\sigma = \{\svec \in S^2\,|\, \sgn (\svec\cdot \jhat) = \sigma_j\}$.
The wrapping numbers $w^\sigma$
may be expressed in terms of $(e,k,\Omega)$, as follows \cite{rz2003,
  mrz2006}:
\begin{equation}
   \label{eq:trapped_area_and_wrapping_number}
 w^\sigma = \fpi \Omega + \half \sum_j \sigma_j k_j + \smallfrac18 e_* (1 -
 8\delta_{\sigma e}),
\end{equation}
where $\delta_{\sigma e}$ equals one if
$\sigma = e$ and is zero otherwise.
Note that (\ref{eq:trapped_area_and_wrapping_number}) implies that
\begin{equation}
  \label{eq:Omega_mod_2pi}
  \Omega =  -2\pi \sum_j \sigma_j k_j -
e_* \pi/2 \mod 4\pi,
\end{equation}
where 
\begin{equation}
  \label{eq:e_*}
  e_* = e_x e_y e_z.
\end{equation}
(\ref{eq:trapped_area_and_wrapping_number}) may be inverted to obtain
$(e,k,\Omega)$ in terms of the wrapping numbers, 
\begin{align}
  e_t &= -\sum_\sigma \sigma_r \sigma_s w^\sigma, \ \text{for $r$, $s$,
  $t$ distinct,}
  \label{eq:e_from_w}\\
  k_r &= \fourth \sum_\sigma \sigma_r w^\sigma + \fourth e_r e_*,   \label{eq:k_from_w}\\
 \Omega &= \sum_\sigma\frac{\pi}{2} w^\sigma .  \label{eq:Omega_from_w}
\end{align}
(Similar relations are described for a general polyhedron in the Appendix.)

A topology for a prism configuration $\nvec$ may be
specified as a set of edge signs, kink numbers and trapped areas for
each vertex.  The conditions for the topology to be admissible (ie,
realisable by a configuration continuous away from the vertices) are
readily expressed in terms of $(e,k,\Omega)$;   
pairs of edge signs associated with a single edge must be compatible,
the absence of surface singularities implies sum
rules for the kink numbers on each face of the prism, and the absence of
interior singularities implies a sum rule for the trapped areas.

We say that an octant topology $(e,k,\Omega)$ is {\it conformal}
(resp.~{\it anticonformal}) if
every nonzero wrapping number is negative (resp.~positive).  
From (\ref{eq:trapped_area_and_wrapping_number}), one can show that
$(e,k,\Omega)$ is conformal
(resp.~anticonformal) if and only if  $\Omega \le -\Omega_-(e,k) $
(resp.~$\Omega \ge \Omega_+(e,k)$), where  
 \begin{align}
    \label{eq:Omega_+-}
    \Omega_{e_*}(e,k) &= 2\pi \sum_j |k_j| + 2\pi
    \begin{cases}
     +\frac74,& \text{if}\ e_* e_j  k_j \le 0\ \text{for all}\ j,\\
     -\frac14,& \text{otherwise}.
    \end{cases},\nonumber\\
    \Omega_{-e_*}(e,k) &= 2\pi \sum_j |k_j| - 2\pi
    \begin{cases}
     +\frac74,& \text{if}\ e_* e_j  k_j > 0\ \text{for all}\ j,\\
     -\frac14,& \text{otherwise}.
    \end{cases}
\end{align}
If $(\Omega,e,k)$ is an octant topology, then $\Omega$ differs
from $\Omega_+(e,k)$ or $\Omega_-(e,k)$ by some multiple of $4\pi$
(ie, some number of whole coverings of the sphere).  If $\Omega =
\Omega_\pm(e,k)$, then $(e,k,\Omega)$ has at least one wrapping number
equal to zero.  
\subsection{Conformal configurations}\label{sec: con}
$\nuvec: O \rightarrow S^2$ can be
represented by a 
complex function $F(w,\wbar)$ in the standard way via the stereographic
projection $S^2 \rightarrow \Cc^*$ ($\Cc^*$ is the extended complex plane),
\begin{equation}
F\left(\frac{s_x + i s_y}{1 + s_z},
\frac{s_x - i s_y}{1 + s_z}\right) = 
\left(
\frac{\nu_x + i \nu_y}{1 + \nu_z}
 \right)
(\svec).  \label{eq:F(w,wbar)}
\end{equation}
The domain of $F$ is 
the positive quarter-unit disk (the image of $O$ under the projection),
\begin{equation}
  \label{eq:Q}
  Q = \{ w \in \Cc |\, 0 \le \Re w \le 1,\ 0 \le \Im w \le 1,\  |w| \le 1\}.
\end{equation}
Letting $w_j(\alpha)$ denote the projections of the parameterised
boundaries $\svec_j(\alpha)$ of $\partial O_j$, we have that
\begin{equation}\label{eq:w_j(alpha)}
    w_x(\alpha) = \tan(\alpha/2), \quad  w_y(\alpha) = i 
   (1-\tan(\alpha/2)),\quad  w_z = e^{i\alpha}.
\end{equation}
A standard calculation gives 
\begin{equation}\label{eq: grad nu squared}
\left(\grad\nuvec\right)^2 = 4 \frac{|F_\wbar|^2 + |F_w|^2}{(1 + |F|^2)^2},
\end{equation}
so that (cf (\ref{eq: E_2 sans a}) and (\ref{eq: E_1 sans a}))
\begin{align}
    E_{(2)}(\nuvec)
&= \int_Q 4 \frac{ |\partial_\wbar F|^2 + |\partial_w F|^2 }{(1 + |F|^2)^2}
    d^2 w,   \label{eq:E in w}\\
 E_{(1)j}(\nuvec) &= \int_{0}^{\pi/2}  4\left(\frac{|\partial_\wbar F|^2 +
  |\partial_w F|^2}
{(1 + |F|^2)^2}\right)(w_j(\alpha))
  \left|\frac{dw_j}{d\alpha}(\alpha)\right|^2 \, d\alpha.  \label{eq:E_1j_general}
\end{align}

Of particular interest are configurations for which $F$ is conformal
(a function of $w$ only) or anticonformal. 
For definiteness we will consider the conformal case, and
write $F(w,\wbar) = f(w)$. If $f$ has a meromorphic extension to the extended complex
plane, then tangent boundary conditions imply that
$f$ is real when $w$ is real, $f$ is imaginary when $w$ is imaginary,
and $|f| = 1$ when $|w| = 1$. It follows that
if $w_*$ is a zero of $f$, then so are
$-w_*$ and $\pm \wbar_*$, while $\pm 1/w_*$ and $\pm
1/\wbar_*$ are poles. Therefore,
$f$ is in fact a rational function of the form
\begin{equation}\label{eq: f form}
 f = \lambda w^n A B C.
\end{equation}
Here, $A$ contains the real zeros and poles of $f$, $B$ the imaginary
zeros and poles, and $C$ the strictly complex zeros and poles. $n$ is
an odd integer giving the order of the zero or pole at the origin, and
$\lambda = \pm 1$ is an overall sign.  $A$, $B$ and $C$ may be written
explicitly as
\begin{subequations}\label{eq: ABC}
\begin{align}
    A(w) &=  \prod_{j=1}^a\left( \frac{w^2 - r_j^2}{r_j^2 w^2 - 1}
    \right)^{\rho_j},\label{eq:A}\\
    B(w) &= \prod_{k=1}^b\left( \frac{w^2 + s_k^2}{s_k^2 w^2 + 1}
    \right)^{\sigma_k},\label{eq:B}\\
    C(w) &= \prod_{l=1}^c\left( \frac{(w^2 - t_l^2)(w^2 - \tbar_l^2)}
      {(t^2_l w^2 - 1)({\tbar}_l^2w^2 - 1)} \right)^{\tau_l}.\label{eq:C}
\end{align}
\end{subequations}
Here, $a$ is the number of real zeros and poles in $Q$ (excluding the origin),
$b$ the number of
imaginary zeros and poles in $Q$ (excluding the origin), and $c$ the number of strictly
complex zeros and poles in $Q$.
$r_j$, with $0 < r_j < 1$, denotes the real zeros ($\rho_j = 1$)
and poles ($\rho_j = -1$); similarly, $is_k$, with $0 <
s_k < 1$ and $\sigma_k = \pm 1$, denote the imaginary zeros and poles, and $t_l$, with $0 < |t_l| < 1$, $\Re t_l, \Im t_l > 0$ and
$\tau_l = \pm 1$ denote the strictly complex zeros and poles.

In terms of these parameters, the edge signs, kink numbers and trapped area of
conformal configurations are given by \cite{mrz2004b,mrz2006}
\begin{equation}
  \label{eq:edge signs}
e_x = \lambda(-1)^a,\
e_y = \lambda(-1)^b(-1)^{(n-1)/2},\ 
e_z = \sgn\, n,
\end{equation}
\begin{align}
  \label{eq:ks}
  k_x &= -\half (-1)^{b} e_y 
\left(
{ \sum_{k = 1}^b} (-1)^{k}\sigma_k + 
\half (1 - (-1)^{b}) e_z\right)
,\nonumber\\
  k_y &= -\half (-1)^a e_x 
\left(
{ \sum_{j = 1}^a} (-1)^{j} \rho_j + \half (1-(-1)^{a})e_z
\right)
,\\
k_z &= \fourth\left(e_xe_y - n\right) 
-\half{ \sum_{j=1}^a} \rho_j
-\half{ \sum_{k=1}^b} \sigma_k
-{ \sum_{l=1}^c} \tau_l,\nonumber
\end{align}
\begin{equation}
  \label{eq:Omega formula}
  \Omega = -\half (|n| + 2(a + b) + 4c)\pi.
\end{equation}
The expressions for the edge signs follow from evaluating $f$ at $1$,
$i$ and $0$, while the formula for $\Omega$ follows from noting that
$8\Omega$ is just
($-4\pi$ times) the degree of $f$, 
which for a  meromorphic function is the number of its zeros 
counted with multiplicity.  The formulas for the kink numbers require
a bit more calculation; details are given in \cite{mrz2006}.

  It is easily checked that a conformal configuration $f(w)$ has a
conformal octant topology ($f$ is
orientation-preserving, which implies that the nonzero 
wrapping numbers are negative).  Therefore, octant topologies which are
neither conformal nor anticonformal, ie $(e,k,\Omega)$ for which
$-\Omega_{-}(e,k) < \Omega < -\Omega_{+}(e,k)$, cannot be realised by
$F =f(w)$ or $F = f(\wbar)$.
In \cite{mrz2006} we establish a converse result, namely that evey conformal
(resp.~anticonformal) octant topology has a conformal
(resp.~anticonformal) representative.  

\subsection{Proof of Lemma \ref{lem:lem.2}}\label{sec: pf of lem2}
\begin{proof}
Given an admissible topology $h$ for $P$, let $(e,k,\Omega)$ denote
the associated octant topology on  $O$ with wrapping numbers
$w^\sigma$.
We construct $\nuvec$, or rather its complex representative
$F(w,\wbar)$, with  topology $(e,k,\Omega)$ in
Step 1.  We establish the estimate
(\ref{eq:lem2_1 sans a}) for $E_{(2)}(\nuvec)$ in Step 2.  We then
show that 
\begin{equation}
  \label{eq:E_1j_est}
  E_{(1)j}(\nuvec) \lesssim 1 + k_j^2
\end{equation}
for $j = z$ (Step 3) and $j = x, y$ (Step 4).  From
(\ref{eq:Omega_mod_2pi}), $\Omega \ne 0$.  (\ref{eq:Omega_from_w})
then implies that the wrapping numbers cannot all vanish, so that, from
(\ref{eq:k_from_w}),
\begin{equation}
  \label{eq:k_ll_sum_w}
\sum_j  1 + |k_j|^2 \lesssim \sum_{\sigma} |w^\sigma|^2.
\end{equation}
The bound  (\ref{eq:lem2_2 sans a}) for $\sum_j E_{(1)j}(\nuvec)$ 
then follows from (\ref{eq:E_1j_est}) and (\ref{eq:k_ll_sum_w}).\\

\noindent {\it Step 1.  Definition of $F$}\quad  
In general, the octant topology $(e,k,\Omega)$ is neither conformal nor anticonformal.  
We  will take $F$ to be conformal outside a small disk in
$Q$ with conformal topology $(e,k, \Omega_*)$.
Inside the disk, $F$ is made to cover the complex plane
$(\Omega - \Omega_*)/(4\pi)$ times; this will ensure
that $F$ has the required topology.

Let 
\begin{equation}
  \label{eq:Omega_*}
  \Omega_* = -2\pi\left(
\sum_j |k_j| + 1 - \smallfrac34 e_*
\right).
\end{equation}
Using (\ref{eq:Omega_+-}), one can check that $\Omega_*$ is equal to
either $-\Omega_-(e,k)$ or $-(\Omega_-(e,k)-4\pi$; in either case,
$(e,k,\Omega_*)$ is a conformal octant topology.  Denoting its
wrapping numbers by $w_*^{\sigma}$, it follows that $w_*^\sigma \le 0$
and, from (\ref{eq:trapped_area_and_wrapping_number}) and
(\ref{eq:Omega_+-}), that at least one of its wrapping numbers either
vanishes or is equal to $-1$;
\begin{equation}
  \label{eq:min_w_*sigma}
  |w_*^{\sigma_0}| \le 1 \quad \text{for some $\sigma_0$.}
\end{equation}

An explicit conformal representative $f(w)$ with octant topology
$(e,k,\Omega_*)$ of the form (\ref{eq: f form}) and (\ref{eq: ABC}) is
obtained by taking
\begin{gather}
  n = (2 - e_*) e_z,\quad \lambda = e_x,\nonumber\\
a = 2|k_y|,\quad \rho_j = -(-1)^{j} e_x \sgn k_y, \quad 1 \le j \le a,\nonumber\\
b = 2|k_x|,\quad \sigma_k = -(-1)^{k} e_y \sgn k_x, \quad 1 \le k \le b,\nonumber\\
c = |k_z| + \half(1-e_*),\quad \tau_l =  
\begin{cases}
  -\sgn k_z, &  l \le |k_z|,\\
 -e_z,&  l = |k_z| + 1,
\end{cases}\label{eq:f}
\end{gather}
as can be verified from
(\ref{eq:edge signs})--(\ref{eq:Omega formula}).
(The reason for introducing $\Omega_*$ --- we could  use
$-\Omega_-(e,k)$ instead  --- is that conformal 
representatives for $(e,k,-\Omega_-(e,k))$ entail several special
cases.)
%
%
Let
\begin{equation}
  \label{eq:M}
  4\pi m = \Omega - \Omega_*(e,k).
\end{equation}
Then, from (\ref{eq:Omega_mod_2pi}), $m$ is an integer, and from
(\ref{eq:trapped_area_and_wrapping_number}), 
\begin{equation}
  \label{eq:w_and_w_*}
  w^\sigma = w_*^{\sigma} + m.
\end{equation}

If $m = 0$, we let
\begin{equation}\label{eq:F_m_0}
  F(w,\wbar) = f(w).
\end{equation}
Otherwise, 
let $w_0$ denote a regular
point of $f$ in the interior of $Q$, and let $D_\epsilon(w_0)$ denote
the open $\epsilon$-disk about $w_0$.  Choose $\epsilon$ sufficiently
small so that $D_{2\epsilon}(w_0)$ is contained in $Q$ and contains no
poles of $f$.  Let
\begin{equation}
  \label{eq:s}
  s(w,\wbar) = \frac{|w-w_0| - \epsilon}{\epsilon}
\end{equation}
(so that $s$ varies between $0$ and $1$ as $|w - w_0|$ varies between
$\epsilon$ and $2\epsilon$). 
Then for $m > 0$ we define
\begin{subequations}\label{eq:F}
\begin{equation}
  \label{eq:F>}
  F(w,\wbar) =
  \begin{cases}
    f(w), & |w - w_0| \ge 2\epsilon,\\
 s 
f(w) + (1-
s)
(f(w_0) + (w - w_0)^{m}),&
    \epsilon < |w - w_0| < 2\epsilon,\\
    f(w_0) + \epsilon^{2m}(\wbar - \wbar_0)^{-m},&  |w - w_0| \le \epsilon,
\end{cases} 
\end{equation}
while for $m < 0$ we define
\begin{equation}
  \label{eq:F<}
  F(w,\wbar) =
  \begin{cases}
    f(w), & |w - w_0| \ge 2\epsilon,\\
 s 
f(w) + (1-
s)
(f(w_0) + (\wbar - \wbar_0)^{-m}),&
    \epsilon < |w - w_0| < 2\epsilon,\\
    f(w_0) + \epsilon^{-2m}(w - w_0)^{m},&  |w - w_0| \le \epsilon.
\end{cases} 
\end{equation}
\end{subequations}

$F$ coincides with $f$ on $\partial Q$, so that $F$ has the
same edge signs and kink numbers as $f$, namely $e$ and $k$. 
Let us verify that $F$ has trapped area $\Omega$.  For $m = 0$ this is automatic.
Otherwise, for definiteness,  suppose that $m > 0$ (the case $m < 0$ is
treated similarly).  From (\ref{eq:trapped_area_1}) and
(\ref{eq:F(w,wbar)}) one can show that
\begin{equation}
  \label{eq:Omega w}
\Omega(F) = 
\int_Q 4 \frac{|\partial_\wbar F|^2 - |\partial_w F|^2
  }{(1 + |F|^2)^2}\, d^2 w.
\end{equation}
Divide the domain of integration
as in (\ref{eq:F}).  
The contribution  from $|w - w_0| > 2\epsilon$
is, to $O(\epsilon^2)$, just the trapped area of
$f$, namely $\Omega_*(e,k)$.
Consider next the contribution from $D_{\epsilon}(w_0)$.  From
(\ref{eq:F}), $F(D_{\epsilon}(w_0))$ covers the extended complex
plane $m$ times with positive orientation, apart from an
$\epsilon^m$-disk about $f(w_0)$.  It follows that its contribution to
the integral in
(\ref{eq:Omega w}) is, to within $O(\epsilon^{2m})$ corrections,
$4\pi m$.  The remaining contribution, from the annulus
$D_{2\epsilon}(w_0) - D_{\epsilon}(w_0)$, is $O(\epsilon^2)$.  This is
because the area of the annulus is $O(\epsilon^2)$, while the
integrand in (\ref{eq:Omega w}) may be bounded independently of
$\epsilon$ (since, by assumption, $f$ has no poles in $D_{2\epsilon}(w_0)$).
Since the trapped area is an odd multiple of $\pi/2$ 
(cf (\ref{eq:Omega_mod_2pi})), it follows that,
for small enough $\epsilon$, $F$ has trapped area $ \Omega_*(e,k) +
4\pi m = \Omega$.

Clearly the
topology of $F$ does not depend on the positions of the zeros and
poles of $f$.  As will be evident in Step 2 below, neither does the
octant energy
$E_{(2)}(\nuvec)$, at least to leading order in $\epsilon$.  However,
the edge energies, $E_{(1)j}(\nuvec)$, do depend on the positions of
the zeros and poles.
As will be evident in Steps 3 and 4, 
to obtain good control of the edge energies, the  $a$ real and $b$ imaginary
zeros and poles in $Q$
should be kept away from the origin, the unit circle and 
each other, while the $c$ strictly complex imaginary zeros and poles should be kept close
to the unit circle and away from the real and imaginary axes and each other.
Anticipating these requirements, in (\ref{eq: ABC}) we take
\begin{equation}\label{eq: zeros and poles}
  r_j = \frac14 + \frac{j}{2a}, \ \ s_k = \frac14 +
  \frac{k}{2b},\ \
t_l = \left(1 - \frac{1}{c+1}\right)^{1/2} e^{i \alpha_l},\ \frac{\alpha_l}{\pi}  = \frac18
  + \frac{l}{4(c+1)}.
\end{equation}
These imply the properties
\begin{subequations}
  \label{eq: zero pole properties}
  \begin{gather}
    1/4 < r_j \le 3/4, \quad r_{j+1} - r_j = 1/(2a), \label{eq: r_j
    props}\\
 1/4 < s_k \le 3/4, \quad s_{k+1} - s_k = 1/(2b), \label{eq: s_k
    props}\\
\pi/8 < \alpha_l \le 3\pi/8, \quad 1 - |t_l|^2  = 1/(c+1) \label{eq: t_l
    props}
  \end{gather}
\end{subequations}
which will be useful in what follows.  We note that, with $r_j$ and
$\rho_j$ as given in (\ref{eq:f}) and (\ref{eq: zeros and poles}), 
the real zeros and poles of $f$
alternate along the interval $(0,1]$; similarly, with $s_k$ and
$\sigma_k$ as given in (\ref{eq:f}) and  (\ref{eq: zeros and poles}),
 the imaginary zeros and poles alternate along $(0,i]$.
\\


\noindent {\it Step 2.  Estimate of $E_{(2)}(\nuvec)$}\quad  
The expression (\ref{eq:E in w}) for  $E_{(2)}(\nuvec)$ differs from
the expression (\ref{eq:Omega w}) for $\Omega$ only in the relative
sign of the $w$- and $\wbar$-derivative terms.  Arguing as for (\ref{eq:Omega w}),
we see that, to order $\epsilon^2$, the
contributions to $E_{(2)}(\nuvec)$ from $Q-D_{2\epsilon}(w_0)$,
$D_{2\epsilon}(w_0) - D_{\epsilon}(w_0)$, and $D_{\epsilon}(w_0)$ are,
respectively, $|\Omega_*|$, $0$ and $4\pi |m|$.  Therefore,
\begin{equation}
  \label{eq:E_2_1}
   E_{(2)}(\nuvec)\lesssim |\Omega_*| + 4\pi |m| = \sum_\sigma |w_*^{\sigma}|
   \frac{\pi}{2} + 4\pi |m|,
\end{equation}
where we have used (\ref{eq:Omega_from_w}) and the fact that
$w_*^{\sigma} \le 0$.
From (\ref{eq:w_and_w_*}), $|w_*^{\sigma}| \le |w^{\sigma}| + |m|$.
Also, from (\ref{eq:min_w_*sigma}) and  (\ref{eq:w_and_w_*}), $|m| =
|w^{\sigma_0} - w^{\sigma_0}_*| \le |w^{\sigma_0}| + 1 \le
2\sum_\sigma |w^\sigma|$.
Substituting these results into
(\ref{eq:E_2_1}), we get 
that 
\begin{equation}
  \label{eq:E_2_11}
   E_{(2)}(\nuvec)\lesssim \sum_\sigma |w^{\sigma}|.
\end{equation}
verifying (\ref{eq:lem2_1 sans a}).\\

\noindent{\it Step 3.  Estimate of $E_{(1)z}(\nuvec)$}\quad  
On
 $\partial O_z$, $F = f$ and $|f| = 1$.  Also, from (\ref{eq:w_j(alpha)}), $|dw_z/d\alpha| =
 1$.  Therefore, from 
 (\ref{eq:E_1j_general}),
\begin{equation}
   \label{eq:int_for_E_1z}
 E_{(1)z}(\nuvec) = \int_0^{\pi/2} \left| \frac{f'}{f}  \right|^2 (e^{i\alpha})\, d\alpha.
\end{equation}
Below we show that 
\begin{equation}
  \label{eq:estimate_of_f'/f}
   \left| \frac{f'}{f} \right|(e^{i\alpha}) \lesssim 1 + |k_z|,
\end{equation}
which then yields the required estimate 
(\ref{eq:E_1j_est})
for $j = z$.

To verify (\ref{eq:estimate_of_f'/f}), we note that, from (\ref{eq: f form}),
\begin{equation}
  \label{eq:log_f}
   \left| \frac{f'}{f} \right| =  |n| + \left| \frac{A'}{A} \right | + 
\left| \frac{B'}{B} \right | + \left| \frac{C'}{C} \right |.
\end{equation}
From the expression (\ref{eq:A}) for $A$ and the positions (\ref{eq: r_j
    props}) of its zeros and poles one calculates that
\begin{equation}
  \label{eq:log_A}
\frac{A'}{A}(w) =   \pm \frac{w}{a} \sum_{J=1}^{a/2} 
\left(
\frac{ r_{2J-1} + r_{2J}}{(w^2 - r_{2J-1}^2) (w^2 - r_{2J}^2)} + 
\frac{ r_{2J-1} + r_{2J}}{((r_{2J-1}^2w^2 - 1) ( r_{2J}^2w^2 - 1)}
\right).
\end{equation}
From (\ref{eq: r_j
    props}), $|r_{2J} + r_{2J-1}| \le 3/2$ while, for $w \in \partial
    O_z$, we have that 
$|w^2 - r_j^2|, |r_j^2 w^2 - 1| \ge 1/16$.
Therefore, 
\begin{equation}
  \label{eq:log_A_2nd}
\left|\frac{A'}{A}\right|(w) \lesssim   \sum_{J=1}^{a/2} \frac{1}{a} \lesssim
1, \quad w \in \partial O_z.
\end{equation}
A similar calculation (cf (\ref{eq:B}) and (\ref{eq: s_k
    props})) shows that
\begin{equation}
  \label{eq:log_B_2nd}
\left|\frac{B'}{B}\right|(w) \lesssim 
1, \quad w \in \partial O_z.
\end{equation}

It remains to estimate $C'/C$.  Without loss of generality, we may
assume that $c > 0$ (otherwise, $C = 1$ and $C' = 0$).  From
(\ref{eq:C}) and (\ref{eq: t_l props}),
\begin{equation}
  \label{eq:log C 1}
  \frac{C'}{C}(e^{i\alpha}) = 2e^{-i\alpha} \frac{1}{c+1} \sum_{l=1}^c \tau_l \left(
\frac{ (1 + |t_l|^2)}{|e^{2i\alpha} - t_l^2|^2} + \frac{ (1 + |t_l|^2)}{|e^{2i\alpha} - \tbar_l^2|^2}
\right).
\end{equation}
Write the denominators (\ref{eq:log C 1}) as
\begin{align}\label{eq: C_denoms_1}
|e^{2i\alpha} - \tbar_l^2|^2  &= (1 - |t_l|^2)^2 + 4|t_l|^2
\sin^2(\alpha_l + \alpha),
\nonumber\\
|e^{2i\alpha} - t_l^2|^2  &= (1 - |t_l|^2)^2 + 4|t_l|^2
\sin^2(\alpha_l - \alpha).
\end{align}
From (\ref{eq: t_l
    props}), one has that $1 - |t_l|^2 = 1/(c+1)$ while, for $0 \le
    \alpha \le \pi/2$, we have that $\pi/8 \le \alpha + \alpha_l \le 7 \pi/8$ and
    $-\pi/2 < \alpha - \alpha_l < \pi/2$, so that
\begin{align}\label{eq: C_denoms_2}
|e^{2i\alpha} - \tbar_l^2|^2  &\ge \sin^2(\pi/8), 
\nonumber\\
|e^{2i\alpha} - t_l^2|^{2}  &
\ge  \frac{1}{(c+1)^2} + \frac{8}{\pi^2} (\alpha_l - \alpha)^2
\end{align}
(we have used $(\pi/2) \sin |x| > |x|$ for $|x| \le \pi/2$).
Substituting (\ref{eq: C_denoms_2}) into (\ref{eq:log C 1}) and
using $\alpha_l = \pi/8 + l\pi/(4(c+1))$
(cf (\ref{eq: zeros and poles})) and $1 + |t_l|^2 < 2$
(cf (\ref{eq: t_l
    props})), we get that
\begin{multline}
  \label{eq:log C 2}
\left|  \frac{C'}{C}(e^{i\alpha})\right| 
\le \frac{2}{c+1}\sum_{l=1}^c \left(\frac{2(c+1)^2}{1 + \half(l -
    (c+1)(4\alpha/\pi - \half))^2}+\frac{2}{\sin^2\pi/8}\right) \\ 
\lesssim (c+1) \sum_{l=0}^c \frac{1}{1 + l^2} + 1 \lesssim
 c + 1.
\end{multline}
We substitute (\ref{eq:log_A_2nd}), (\ref{eq:log_B_2nd}) and
(\ref{eq:log C 2}) into (\ref{eq:log_f}) to get that $|f'/f| \lesssim c + 1 +
|n|$ on $\partial O_z$.
Since $c$ 
is equal to $|k_z|$ or $|k_z| + 1$ and  $|n| \le
3$ (cf (\ref{eq:f})), 
the required estimate (\ref{eq:estimate_of_f'/f}) follows.\\

\noindent{\it Step 4.  Estimate of $E_{(1)x}(\nuvec)$ and
  $E_{(1)y}(\nuvec)$ }\quad 
We establish (\ref{eq:E_1j_est}) for $j = x$ and $j = y$.
For definiteness, we consider
$ E_{(1)x}(\nuvec)$, and show that
\begin{equation} \label{eq:E_1_x} 
   E_{(1)x}(\nuvec) \lesssim 1 + |k_x|^2
\end{equation}
(the calculations for  
$E_{(1)y}(\nuvec)$ are essentially the same).
On $\partial O_x$, $F =  f$,
$f$ is real and, from (\ref{eq:w_j(alpha)}), $dw_x/d\alpha = \half
\sec^2\alpha/2 \le 1$ for $0 \le \alpha \le \pi/2$.
It will be convenient to
 parameterise $\partial O_x$ by $0 \le w \le 1$ rather than by $\alpha$.  
(\ref{eq:w_j(alpha)}) and (\ref{eq:E_1j_general}) give that
 \begin{equation}
   \label{eq:E_1_x_2}
   E_{(1)x}(\nuvec) = 
 4 \int_0^1 \frac{{f'}^2}{(1 +
   f^2)^2}\left|\frac{dw_x}{d\alpha}\right|\,dw \le 
4  \int_0^1 \frac{{f'}^2}{(1 +
   f^2)^2}\,dw.
 \end{equation}
 
 The estimate  (\ref{eq:E_1_x}) requires more calculation than the corresponding result
 for $j = z$.
It turns out that a pointwise
 bound on the integrand in (\ref{eq:E_1_x_2}) is not sufficient, as
 $f'^2/(1+f^2)^2 \gg |k_x|^3$ on $\partial O_x$.  The
 domain on which $f'^2/(1+f^2)^2 >>  |k_x|^3$ has measure of
 order $1/|k_x|$, in keeping with (\ref{eq:E_1_x}).  But it will be
 necessary to estimate the integral in (\ref{eq:E_1_x_2}) itself.  We
 note in passing that the complex representation (\ref{eq:F(w,wbar)})
 does not incorporate the cubic symmetries of the octant $O$ in a simple
 way.  Projecting along the axis
 $(1,1,1)/\sqrt{3}$ rather than $\zhat$ would treat the boundaries
 symmetrically,  
and might lead to a simplification of the
 calculations below.

To proceed, we collect the zeros and poles of $f$ along $\partial O_x$ into a
factor $q$ (see (\ref{eq:q}) below for its explicit expression), writing
\begin{equation}
  \label{eq:f=pq}
  f = pq,
\end{equation}
where 
\begin{equation}
  \label{eq:p}
  p = \Abar B C = \lambda \prod_{j=1}^a \left(  \frac{w + r_j}{r_j^2 w^2 - 1}
    \right)^{\rho_j} B C
\end{equation}
has no zeros or poles 
for $0\le w\le 1$.  
We have that 
\begin{multline}
  \label{eq:estimate 1}
  \left|\frac{f'}{1+f^2}\right|  = \left|\frac{p'}{p(pq + 1/(pq))} + p \frac {q'}{1 + p^2 q^2}\right|
\le \frac12 \left|\frac{p'}{p}\right| +
  \max\left(|p|,\frac{1}{|p|}\right) \left|\frac{q'}{1 + q^2}\right|,
\end{multline}
since $|pq| + 1/|pq| \ge 2$, and
\begin{equation}
  \label{eq:q^2p^2}
  1 + p^2q^2 \ge 
  \begin{cases}
     1 + q^2,& |p| \ge 1,\\
p^2(1 + q^2),& |p| \le 1.
  \end{cases}
\end{equation}
With calculations similar to those in Step 3 (details are omitted),
one 
shows that
$|(\log p)'|$ is bounded on $\partial O_x$ independently of 
$k_j$, ie
\begin{equation}
  \label{eq:logp'bound}
  \left| \frac{p'}{p} \right | \lesssim 1, \quad 0 \le w \le 1.
\end{equation}
From (\ref{eq: ABC}) and (\ref{eq:p}), if $a = 0$ then $p(0) = \pm 1$,
ie $\log |p(0)| = 0$.  If
$a \ne 0$,  using (\ref{eq: zeros and poles})
we get that
\begin{multline}
  \label{eq:p(0)}
  \left|(\log |p(0)|)\right| = \sum_{J = 1}^{a/2}
  \left| \log \frac{r_{2J-1}}{r_{2J}}\right|  = 
\sum_{J = 1}^{a/2} 
\left|
\log(1 - 2/(a + 4J))
\right|\\
<  \sum_{J = 1}^{a/2} 4/(a + 4J) <  \int_0^{2a} dx/ (a + x) \le \ln 3,
\end{multline}
so that, in general,  $|(\log |p(0)|)| \lesssim 1$.  Then (\ref{eq:logp'bound}) and (\ref{eq:p(0)}) imply that, for $0 \le w \le 1$,
$|(\log |p(w)| )| \le |\log(|p(0)|)| + \int_0^w |p'/p|\,dw'  \lesssim 1$, or
\begin{equation}
  \label{eq:log_p_bound}
  \max(|p|,1/|p|) \lesssim 1,  \quad 0 \le w \le 1.
\end{equation}
Substituting (\ref{eq:log_p_bound}) and (\ref{eq:logp'bound}) into
(\ref{eq:estimate 1}), we get that
\begin{equation}
  \label{eq:f_estimate 2}
\left|  \frac{f'}{1+f^2} \right|\lesssim  1 
+ \left|\frac{q'}{1 + q^2}\right|.
\end{equation}

The explicit form of $q$ is obtained from 
(\ref{eq: f form}), (\ref{eq:A}) and (\ref{eq:f=pq}), 
\begin{equation}
  \label{eq:q}
  q =  w^n\prod_{j=1}^a (w - r_j)^{\rho_j}.
\end{equation}
We partition $\partial O_x$ into intervals whose endpoints are the zeros and poles of
$q$ and estimate $q'/(1+q^2)$ on each. We consider in detail the interval
$I_J = (r_{2J-1},r_{2J})$; the other intervals are treated similarly.
Let $x = 2a(w - r_{2J-1})$, 
so that $x$ varies between $0$ and $1$ on $I_J$.  On $I_J$, we 
write
\begin{equation}
  \label{eq:q_J}
  q(w(x)) =  w^n(x) \left(g_J(x) h(x)\right)^{\rho_{2J-1}},
\end{equation}
where
\begin{equation}
  \label{eq:h}
  h(x) = \frac{x}{x-1}
\end{equation}
contains the zero and pole at the endpoints of $I_J$ (an explicit
expression for $g_J(x)$ is given in (\ref{eq:g}) below).
It is straightforward to show (the calculation is similar to that
in (\ref{eq:estimate 1})) that 
\begin{equation}
  \label{eq:q'_estimate}
   \left|\frac{q'}{1 + q^2}(w(x))\right|  \le \half |n| +
   \left(\frac{dw}{dx}\right)^{-1} 
\left(\half \left|\frac{dg_J/dx}{g_J}\right| +
   \left|\frac{g_J dh/dx}{1 + g_J^2 h^2}\right|\right). 
\end{equation}
Substituting (\ref{eq:q'_estimate}) 
into (\ref{eq:f_estimate 2}) and noting that $|n|\le 3$,
$dw/dx = 1/(2a)$ and $dh/dx = -1/(1-x)^2$, we get that
\begin{equation}
  \label{eq:f_estimate 22}
  \frac{{f'}^2}{(1+f^2)^2} \lesssim  1 
+ a^2 \left|\frac{dg_J/dx}{{g_J}}\right|^2 + 
a^2 \frac{1}{\left((1-x)^2/|g_J| +  x^2 |g_J|\right)^{2}}.
\end{equation}

The integral of the last term in (\ref{eq:f_estimate 22}) may be
estimated as follows:
\begin{equation}
  \label{eq:integral_estimate}
  \int_0^1  
\left((1-x)^2/|g_J| +  x^2 |g_J|\right)^{-2}\,dx
  \lesssim G_J + \Gcal_J^{-1}, 
\end{equation}
where
\begin{equation}
  \label{eq:G_J}
  \Gcal_J =  \min_{x \in [0,1]} |g_J|, \quad  G_J = \max_{x \in [0,1]}
  |g_J|.
\end{equation}
To get (\ref{eq:integral_estimate}), suppose first that
 $\Gcal_J < 1 < G_J$.  We divide $[0,1]$ into the three subintervals 
\begin{equation}
  \label{eq:intervals}
  K = [0,\half G_J^{-1}], \quad L = [\half G_J^{-1}, 1-
  \half\Gcal_J], 
\quad M = [1-\half \Gcal_J,1].
\end{equation}
On $K$, the integrand in (\ref{eq:integral_estimate}) is bounded by
$G_J^2(1-x)^{-4}$, so that the contribution from $K$ to the integral
is dominated by $G_J$.  On $L$, the integrand is bounded by
$(2(1-x)x)^{-2}$ (since, in general, $|a/g| + |bg| \ge 2|ab|^{1/2}$), so the
contribution from $L$ is dominated by $G_J + \Gcal_J^{-1}$.  On $M$,
the integrand is bounded by $g_J^{-2} x^{-4}$, so its contribution is
dominated by $\Gcal_J^{-1}$.  In case $G_J < 1$, let $K = [0,\half]$
and $L = [\half, 1- \Gcal_J/2]$; in case $\Gcal_J > 1$, let $L =
[\half G_J^{-1}, \half]$ and $M = [\half,1]$.

From (\ref{eq:f_estimate 22}) and (\ref{eq:integral_estimate}) it
follows that
\begin{equation}
  \label{eq:f_estimate 4}
\int_{r_{2J-1}}^{r_{2J}}  \frac{{f'}^2}{(1+f^2)^2} dw \lesssim 
  1
+ a \max_{0\le x \le 1} \left|\frac{dg_J/dx}{{g_J}}\right|^2
  + a   (G_J + \Gcal_J^{-1}).
\end{equation}
We need to estimate the terms involving $g_J$. 
From (\ref{eq:q}) and (\ref{eq:q_J}), $g_J$ is given by
\begin{equation}
  \label{eq:g}
  g_J(x) = P_{J-1}^{-1}(x) P_{a/2 - J}(1-x),
\end{equation}
where
\begin{equation}
  \label{eq:P(x)}
  P_N(x) = \prod_{K=1}^N \frac{x + 2K-1}{x+2K} =
  \frac{\Gamma(N+ \half x+\half)}{\Gamma(N + \half x+1)}
\frac{\Gamma(\half x+1)}{\Gamma(\half x+\half )}.
\end{equation}
 Then leading-order asymptotics for $\Gamma(z)$, ie $\log \Gamma(z) \sim
 (z-\half)\log z - z$ and $(\log \Gamma)'(z) \sim \ln z$
(see, eg, \cite{abramowitz}), yields
\begin{equation}
  \label{eq:P_asymp}
   P_N(x) \lesssim N^{-1/2} \ \text{and} \ \left|\frac{P_N'}{P_N}\right|(x) \lesssim 1
\  \text{ for $0 \le x \le 1$},
\end{equation}
which in turn imply the estimates
\begin{align}
\left|
\frac{dg_J/dx}{g_J}
\right| &\lesssim 1, \quad 0 \le x \le 1,\nonumber\\
 G_J, \Gcal_J^{-1} &\lesssim 
 \begin{cases}
((a/2 + 1 - J)/J)^{1/2}, &1 \le J < a/4,\\
(J/(a/2 + 1 - J))^{1/2}, &a/4 \le J \le a/2.
 \end{cases}
 \label{eq:g_J_estimates}
\end{align}
Substituting (\ref{eq:g_J_estimates}) into (\ref{eq:f_estimate 4}), we get that
\begin{equation}
  \label{eq:f_estimate 5}
\int_{r_{2J-1}}^{r_{2J}}  \frac{{f'}^2}{(1+f^2)^2} dw \lesssim 
  1 + a 
+ a 
\left(
\frac{a/2 +1- J}{J}\right)^{1/2} +
a \left(
\frac{J}{a/2 +1- J}\right)^{1/2}.
\end{equation}

Estimating the contribution from the interval $[r_{2J}, r_{2J+1}]$ to
the integral 
(\ref{eq:f_estimate 22}) is carried out
in much the same way.  The differences are that i) $h(x)$ is replaced
by $1/h(x) = -h(1-x)$ (which may be accommodated by the substitution $x
\rightarrow 1-x$) and ii) $g_J(x)$ in (\ref{eq:g}) is
replaced by 
\begin{equation}\label{eq: g_J alt}
   \frac{x + 2J-1}{x + 2J - a} P_{J-1}(x) P_{a/2 - J -1}^{-1}(1-x)
\end{equation}
But the expression in (\ref{eq: g_J alt}) and its logarithmic derivative satisfy the same
bounds as do $g_J$ and $dg_J/dx$ in (\ref{eq:g_J_estimates}).  Thus, 
the integral of $f'^2/(1+f^2)^2$ over  $[r_{2J}, r_{2J+1}]$ satisfies
the same bound
(\ref{eq:f_estimate 5}) as does the integral over $[r_{2J-1},
r_{2J}]$. We obtain a bound on the collective contribution from the intervals
$[r_j,r_{j+1}]$ by summing over $J$ in (\ref{eq:f_estimate 5}),
\begin{equation}
  \label{eq:f_estimate_5}
  \int_{r_1}^{r_a}  \frac{{f'}^2}{(1+f^2)^2} dw \lesssim 1 + a^2  +
  a\sum_{J = 1}^{a/2} \left(\frac{a/2 + 1 - J}{J}\right)^{1/2} +
 a \sum_{J = 1}^{a/2} \left(\frac{J}{a/2 +1- J}\right)^{1/2}.
\end{equation}
The first sum may be estimated as 
\begin{equation}
  \label{eq:sum_estimate}
  \sum_{J = 1}^{a/2} \left(\frac{a/2+1 -J}{J}
\right)^{1/2} \lesssim \int_0^{a/2} \left(\frac{a/2 - y}{y}
\right)^{1/2}dy = \frac{a}{2} \int_0^1
\left(\frac{1-s}{s}\right)^{1/2}ds = \frac{\pi}{4} a,
\end{equation}
and the second is similarly bounded.
Thus
\begin{equation}
  \label{eq:f_estimate_6}
  \int_{r_1}^{r_a}  \frac{{f'}^2}{(1+f^2)^2} dw \lesssim 1 + a^2.
\end{equation}

The contributions from the remaining intervals $[0,r_1]$ and $[r_a,1]$
are treated similarly, and we get
\begin{align}
  \int_{0}^{r_1}  \frac{{f'}^2}{(1+f^2)^2} dw  &\lesssim 1 + a^2,
  \label{eq:f_estimate_7}\\  
\int_{r_a}^{1}  \frac{{f'}^2}{(1+f^2)^2} dw  &\lesssim 1 + a^2 \label{eq:f_estimate_8}.
\end{align}
We give an argument for (\ref{eq:f_estimate_7})
((\ref{eq:f_estimate_8}) is 
treated similarly).
For definiteness, let us assume that $\rho_1 = 1$ (the case
$\rho_1 = -1$ is treated similarly).  On $[0,r_1]$ we write $q = uv$,
where $v = w^n (w-r_1)$ contains the zeros at the endpoints and 
\begin{equation}
  \label{eq:g_and_h_0_1/4}
  u = (w - r_a)^{-1} P_{a/2-1}^{-1}(1 - 2a( w - \fourth))
\end{equation}
contains the remaining factors.
Arguing as in (\ref{eq:q'_estimate}) and (\ref{eq:f_estimate 2}) (but noting that $u$ and $v$ are
functions of $w$,  not a rescaled coordinate $x$), we get that
\begin{equation}
  \label{eq:q'/q_est_0_1/4}
   \frac{f'^2}{(1 + f^2)^2}  \lesssim 1 + 
 \frac{u'^2}{u^2} +
   \frac{u^2 v'^2}{(1 + u^2 v^2)^2 }.
\end{equation}
Clearly $v'^2 \lesssim 1$, so that $ u^2 v'^2/(1 + u^2 v^2)^2 \lesssim u^2$.  From (\ref{eq:P_asymp}) and
(\ref{eq:g_and_h_0_1/4}), $|u'/u|^2 \lesssim a^2$ and $|u|^2 \lesssim a$.
Therefore, $f'^2/(1+f^2)^2 \lesssim 1 + a^2$, and (\ref{eq:f_estimate_8})
follows.  The required bound on $E_{(1)x}(\nuvec)$, (\ref{eq:E_1_x}), follows from
substituting the estimates
(\ref{eq:f_estimate_6})--~(\ref{eq:f_estimate_8}) into the formula (\ref{eq:E_1_x_2}).


\end{proof}

\noindent{\it Remark.} To extend the upper bound of Theorem~\ref{sec: upper bound} to, say, a
general convex polyhedron, one would like to have a generalisation of
the octant configurations of Section~\ref{sec: con}.  These would be
conformal maps, perhaps with singularities, of a general convex
geodesic polygon $\Sigma \subset S^2$ into $S^2$ such that each edge
of $\Sigma$ is mapped into the geodesic which contains it.  \\

\noindent{\it Acknowledgements.} We thank CJ Newton and A Geisow for stimulating our
interest in this area.
AM was partially supported by an EPSRC/Hewlett-Packard Industrial CASE
Studentship.  AM and MZ were partially supported by 
EPSRC grant EP/C519620/1.




\appendix

\section{Wrapping numbers as complete invariants}\label{sec:app}
In \cite{rz2003} we gave a homotopy classification of tangent
unit-vector fields on a convex polyhedron $P$ in terms of a set of
invariants called edges signs, kink numbers and trapped areas.  Here
we show that these invariants can be determined from the wrapping
numbers, so that the wrapping numbers $w^{a\sigma}$ constitute a
complete set of invariants.

It suffices to consider the invariants associated with a single
vertex.  Let $f$ be the number of faces of $P$.  Let $\vvec$ denote a
vertex of $P$, and let $w^\sigma$ denote the wrapping numbers on a
cleaved surface around $\vvec$ (we suppress the vertex label $a$),
where $\sigma$ is an $f$-tuple of signs.  Suppose $\vvec$ has $b \ge
3$ coincident faces and therefore $b$ coincident edges.  Let $E^r$, $1
\le r \le b$, denote the edges coincident at $\vvec$, ordered
consecutively clockwise with respect to a ray from $\vvec$ through the
interior of $P$.  By convention let $E^{b+1} = E^1$.  Let $\Evec^r$
denote the unit vector along $E^r$ directed away from $\vvec$.  Let
$F^r$ denote the face with edges $E^r$ and $E^{r+1}$.  An
(unnormalised) outward normal on $F^r$ is given by
\begin{equation}
  \label{eq:Fvec_j}
  \Fvec^r = \Evec^{r+1}\times\Evec^r.
\end{equation}

Let us explain briefly how the edge signs, kink numbers and trapped areas
are defined (see \cite{rz2003} for details).
Fix a homotopy class $\Ccal(h)$, and let $\nvec \in \Ccal(h)$
denote a representative. The edge sign $e^r$ is given by
the orientation of $\nvec$ along the edge $E^r$ relative to
$\Evec^r$, so that $\nvec(\rvec) = e^r \Evec^r$ for $\rvec \in E^r$. The
kink number $k^r$ is an integer giving the winding number of $\nvec$
about $\Fvec^r$ along a path on the 
face $F^r$ starting on the edge $E^{r+1}$ and ending on the edge $E^r$ 
(tangent boundary conditions imply that $\nvec$ is orthogonal to
$\Fvec^r$ along such path).  A minimal winding (eg, $\nvec$ taking
values along the shortest arc from $\Evec^{r+1}$ to $\Evec^r$
on $S^2$) has kink number equal to zero.  Finally, the trapped area $\Omega$ is
the area in $S^2$ of the image under $\nvec$ of a cleaved surface
about $\vvec$ (the area is normalised so that $S^2$ has area $4\pi$).

It is straightforward to derive the following expression for the
wrapping numbers in terms of $e^r$, $k^r$ and $\Omega$
\cite{rz2003}:
\begin{multline}\label{eq: classnotes}
\Omega = 
4\pi w^{\sigma(\svec)}  +  2\pi\sum_{r=1}^b
\sgn(\Fvec^r\cdot\svec) k^r\\
+  \sum_{r=2}^{b-1} \left(A(e^1
  \Evec^1,e^r\Evec^{r},e^{r+1}\Evec^{r+1}) - 
4\pi  \tau(\svec; e^1 \Evec^{1},e^r\Evec^{r},e^{r+1}\Evec^{r+1})\right).
\end{multline}
Here, $\svec \in S^2$ may be taken to be any 
unit vector which is transverse to every
face of $P$ (including faces which are not coincident at $\vvec$).
$\sigma(\svec)$ gives the sector to which $\svec$ belongs.  For
$\avec, \bvec, \cvec \in S^2$, the quantities
$A(\avec,\bvec,\cvec)$ and $\tau(\svec;\avec,\bvec,\cvec)$ are defined
as follows.  Let $K\subset S^2$ denote the spherical triangle with
vertices $\avec$ , $\bvec$, and $\cvec$ ($K$ is well defined provided
$\avec$, $\bvec$ and $\cvec$ are not coplanar and no pair of them are 
antipodal).  Then
$A(\avec,\bvec,\cvec)$ is the oriented area of $K$ with values
between $-\pi$ and $\pi$ (the sign is given
by $\sgn 
(\avec\cdot (\bvec\times\cvec))$).  The quantity $\tau(\svec;\avec,\bvec,\cvec)$ is given
by
\begin{equation}
  \label{eq:tau}
  \tau(\svec; \avec,\bvec,\cvec) =
  \begin{cases}
    \sgn (\avec\cdot (\bvec\times\cvec)), & \svec \in K,\\
    0,&\svec \notin K.
  \end{cases}
\end{equation}
That is, $ \tau(\svec; \avec,\bvec,\cvec)$ is equal to zero unless
$\svec$ belongs to $K$, 
in which case it is equal to $\pm 1$ according to whether $K$ has
positive or negative area.
Note that
$\tau(\svec;\avec,\bvec,\cvec)$ is well defined if $\svec$ is
transverse to the planes spanned by $\avec, \bvec, \cvec$ taken
pairwise.
%

Our task here is to show that, given the wrapping numbers $w^\sigma$
for topology $h$, we can determine the edge signs, kink numbers and
trapped areas.  We begin by determining the edge signs, specifically $e^r$ and
$e^{r+1}$.  Without loss of generality, we can assume that $r\ne 1$
and $r\ne b$
(note that (\ref{eq: classnotes}) remains valid if the edge indices are
cyclically permuted; there is nothing special about the edge $E^1$).

Let $S\subset S^2$ denote the great circle containing $\Evec^r$ and
$\Evec^{r+1}$.  The four points $\pm \Evec^r$, $\pm \Evec^{r+1}$
partition $S$ into four disjoint open arcs.  Denote these  by $S_m$,
$m = 1, 2, 3, 4$ (the ordering is not important).  
Let $S_{m_*}$ denote the arc 
whose endpoints are $e^r
\Evec^r$ and $e^{r+1}\Evec^{r+1}$.  We determine $m_*$, and hence 
$e^r$ and $e^{r+1}$, by means of the following calculation.
For each $m$, choose some $\svec_m \in S_m$,
and let
\begin{equation}
  \label{eq:svec^pm}
  \svec^{\pm}_m = \svec_m \pm \epsilon \Fvec^r
\end{equation}
(recall that $\Fvec^r$ is normal to $S$) with $\epsilon>0$ small enough so
that $\svec^{\pm}_m$ is transverse to every face of $P$ and so that
\begin{equation}
  \label{eq:svec_prop}
 \sgn( \Fvec^s \cdot \svec^{+}_m) = \sgn(\Fvec^s \cdot \svec^-_m), \quad s \ne r.
\end{equation}

We subtract the two equations obtained by letting $\svec = \svec^\pm_m$ in
(\ref{eq: classnotes}) to obtain
\begin{multline}
  \label{eq:tau_diff}
 w^{\sigma(\svec_m^+)} - w^{\sigma(\svec_m^-)} =\\ -k^r + 
  \tau(\svec^+_m; e^1 \Evec^{1},e^r\Evec^{r},e^{r+1}\Evec^{r+1}) -  
\tau(\svec^-_m; e^1 \Evec^{1},e^r\Evec^{r},e^{r+1}\Evec^{r+1}) .
\end{multline}
Let $K \subset S^2$ denote the spherical triangle with vertices
$ e^1 \Evec^{1}$, $e^r\Evec^{r}$ and $e^{r+1}\Evec^{r+1}$.
If $m \ne m_*$, then neither $\svec^+_m$ nor $\svec^-_m$ lies in $K$,
so that, by (\ref{eq:tau}),
both of the $\tau$-terms in (\ref{eq:tau_diff}) vanish.
On the other hand, if $m = m_*$, then either $\svec^+_{m_*}$ or $\svec^-_{m_*}$ lies in
$K$ but not both, 
so that one of the $\tau$-terms in (\ref{eq:tau_diff}) vanishes
while the other is equal to $\pm 1$.  
Therefore, amongst the four possible values of
$w^{\sigma(\svec_m^+)} - w^{\sigma(\svec_m^-)}$, three will have the
same value and one will be different.  $m_*$ is identified as
the index for which $w^{\sigma(\svec_m^+)} -
w^{\sigma(\svec_m^-)}$ has the different  value.

Once the edge signs are determined, the kink numbers can be obtained
from  (\ref{eq:tau_diff}), and hence the trapped area from (\ref{eq: classnotes}).

\bibliography{lc06}
\end{document}